\documentclass[%
reprint,
nofootinbib,
 amsmath,amssymb,
 aps,
prx,
english,
]{revtex4-2}
\usepackage{graphicx}% Include figure files
\usepackage{dcolumn}% Align table columns on decimal point
\usepackage{bm}% bold math
\usepackage[colorlinks,allcolors=cyan]{hyperref}
\usepackage[capitalise]{cleveref}
\usepackage{soul}
\usepackage{xcolor}
\usepackage{verbatim}
\usepackage[bottom]{footmisc}

\begin{document}

\preprint{APS/123-QED}

\title{Can the chemical potential and the many-body correlated phases of a two-dimensional dipolar fluid be approximated analytically? A Quantum Monte-Carlo study of a universal behavior of dipolar excitons}

\author{Yotam Mazuz-Harpaz}
\altaffiliation{These authors contributed equally to this work.}
\affiliation{Racah Institute of Physics, The Hebrew University of Jerusalem, Jerusalem 9190401, Israel}
\author{Michal Zimmerman}
\altaffiliation{These authors contributed equally to this work.}
\affiliation{Racah Institute of Physics, The Hebrew University of Jerusalem, Jerusalem 9190401, Israel}
\author{Snir Gazit}%
\author{Ronen Rapaport} \email{ronen.rapaport@huji.ac.il}
    \affiliation{Racah Institute of Physics, The Hebrew University of Jerusalem, Jerusalem 9190401, Israel}
\date{\today}

\begin{abstract}

A central challenge in quantum many-body physics is to connect microscopic interactions to experimentally accessible macroscopic states of matter, where correlations extend across multiple length scales. Using quantum Monte Carlo (QMC) simulations of two-dimensional repulsive dipolar bosons, recently realized across various experimental platforms, we show that correlations within the dipolar fluid lead to an unexpected simplification. Specifically, we find that these correlations induce a pronounced depletion region around each particle that suppresses short-range physics and yields an emergent universality, where the fluid is governed by dipolar interactions rather than material-specific details.
Exploiting this universality, we establish a unified, quantitatively validated analytical framework for the dipolar fluid, taking indirect (interlayer) excitons as a model system. We show that analytical interaction energies derived in an earlier study agree with exact QMC results across three correlation regimes: a classical gas, a quantum gas, and a high-density liquid. In contrast, the widely used uncorrelated plate capacitor formula severely overestimates the numerical values.
Building on these expressions, we show that the chemical potential is accurately captured by a three-component decomposition: an ideal Bose-gas kinetic term, a regime-dependent dipolar interaction term, and a high-density quantum confinement term.
Finally, we address a key experimental challenge: extracting exciton densities from measured energy shifts. Our QMC-benchmarked calibration curves for typical gallium arsenide (GaAs) and transition metal dichalcogenide (TMD) heterostructures offer a reliable tool for experimentalists to determine both the fluid's density and collective state. This universal framework extends to other two-dimensional systems, including ultracold dipolar atoms and polar molecules.

\end{abstract}

\maketitle

%\tableofcontents

\section{Introduction}

Quantum fluids with dipole-dipole interactions have emerged as a rich platform for exploring many-body physics that extends beyond the short-range-interaction paradigm. The long-range and anisotropic nature of dipolar forces gives rise to qualitatively new collective behavior, enabling the realization of exotic states across a wide range of experimental platforms. These include quantum droplets and supersolids in cold magnetic atomic gases \cite{Chomaz_dipolar_2023, Bottcher2021New,Chomaz2016Quantum,Ferrier2016Observation,Chomaz2019Long,Bottcher2019Transient,Tanzi2019Observation,Casotti2024Observation,Biagioni2024Measurements}, as well as correlated excitonic phases, such as dark condensates, dipolar-quadrupolar transitions, attractive dipolar fluids, ordered dipolar lattices, insulating phases, and moir\'e physics in semiconductor heterostructures such as GaAs, and in van der Waals heterolayers \cite{Deng_review_2024,Harpaz_dynamical_2019,Slobodkin_2020,Zimmerman2026Quadrupolar,Li_2023,Bai_2023,Lian_2023,Xie_2023,Xie_2024,Meng_2025, hubert_attractive_2019, zimmerman_collective_2022,Mak2021EI,Mak2022EI,Watanabe2022EI,Mak2020Mott,Watanabe2020Mott,Imamoglu2020Mott,Wang2020Mott,Gao_2024_excitonic,Seyler_2019_signatures}. Beyond these specific fields of research, dipolar systems are being actively explored as building blocks for quantum simulation, where their extended interactions enable access to lattice models with significant beyond-on-site coupling terms \cite{Cornish2024Quantum,Su2023Dipolar,Baier2026Extended,Lagoin2026Bose,Lagoin2022Extended}.

At the same time, theoretically describing these dipolar fluids poses a two-fold challenge. First, the long-range nature of dipolar interactions induces strong spatial correlations and places the fluid in the strongly correlated many-body regime
\cite{Astrakharchik2007Quantum,lozovik_strong_2007,Buchler2007Strongly}, where standard perturbative and mean-field techniques generally fail to provide a valid description. In addition, the short-range physics of the fluid is highly system-dependent, governed by exchange and contact interactions and by the microscopic structures of the constituent particles, which differ significantly between atomic, molecular, and solid-state realizations. As a result, it is difficult to construct a simple and universal description of the dipolar system that accurately predicts its collective behavior and connects it to experimentally measured observables.

A notable simplification was proposed by Laikhtman and Rapaport \cite{laikhtman_correlations_2009,laikhtman_exciton_2009}, who argued that in two-dimensional dipolar systems with sufficiently large dipole moments, the repulsive interaction creates a pronounced depletion region around each particle. This depletion suppresses close-range encounters and effectively quenches short-range physics. As a consequence, the system acquires emergent universality: its behavior is governed primarily by dipolar interactions and is largely independent of microscopic details. This mechanism has been directly linked to experimental observations, such as the stabilization of dark-spin Bose–Einstein condensates of dipolar excitons, where short-range spin-exchange processes are strongly suppressed \cite{shilo_particle_2013,cohen_dark_2016,Harpaz_dynamical_2019,Misra_2022,Beian_2017,Combescot_2007}.

Building on this idea, an analytical framework was developed in Refs.~\cite{laikhtman_correlations_2009,laikhtman_exciton_2009} to describe the different correlation regimes of the dipolar fluid. By analyzing the relevant energy and length scales of the system, this framework identifies these distinct correlation regimes as a function of density and temperature and provides explicit \textit{analytical} expressions for the two-body correlation function and the interaction energy in each regime. The resulting correlation phase map offers an appealing and intuitive picture of dipolar many-body physics. However, these results rely on simplifying assumptions and scaling arguments, and their quantitative accuracy has remained largely untested.

\begin{figure} 
\includegraphics[width=0.48\textwidth]{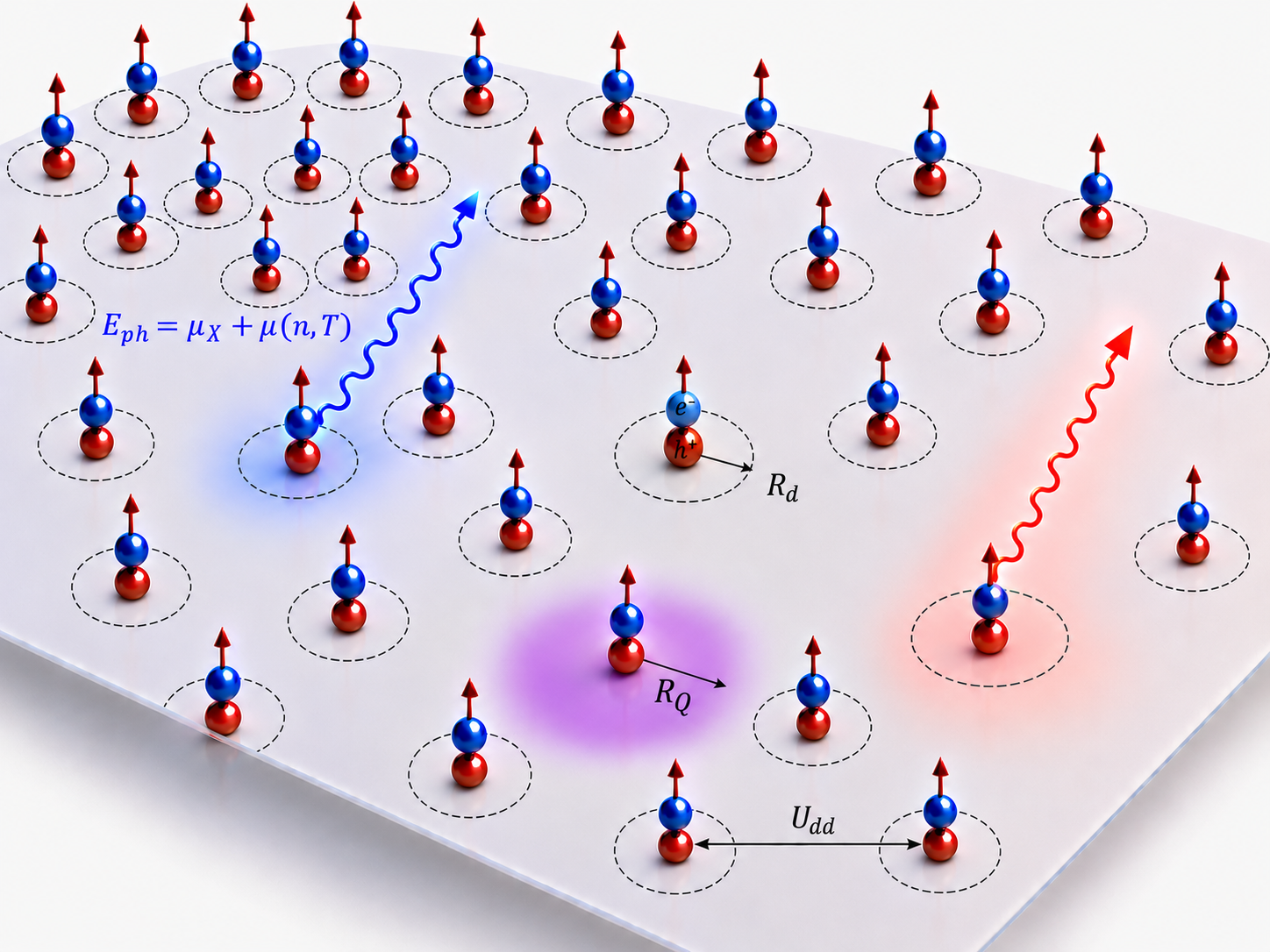}
\centering
\caption{
Schematic illustration of a two-dimensional fluid of dipolar excitons, modeled as point dipolar bosons with parallel dipole moments
oriented perpendicular to the plane. Each exciton is represented by a
spatially separated electron ($e^{-}$, blue) and hole ($h^{+}$, red). The repulsive dipole--dipole interaction between exciton pairs is denoted by $U_{dd}$. The dashed circles represent projections of the depletion regions surrounding the excitons. The depletion radius $R_d$, defined in the text, generally depends on both density and temperature. In the high-density correlated regime (liquid), repulsive interactions within the fluid generate an effective confining
potential for each exciton, giving rise to zero-point positional quantum
fluctuations. The characteristic scale of these fluctuations, $R_Q$, is schematically represented by the purple distribution. Experimentally, these many-body correlations are probed via the emitted exciton energy shifts. When an exciton radiatively recombines, it produces a photon with energy $E_{\mathrm{ph}}=\mu_X+\mu(n,T)$, where $\mu_X$ is the single-exciton emission energy and $\mu(n,T)$ is the density- and temperature-dependent many-body chemical potential of the dipolar fluid. The differently colored photons schematically indicate large energy shifts induced by strong dipolar repulsion in a denser region (blue), and smaller energy shifts in a more dilute region (red).  (Note: The base 3D illustration for this figure was generated using ChatGPT, with mathematical annotations and symbols subsequently added by the authors.)
}
\label{fig:illustration}
\end{figure}

A specific platform for which testing this theoretical framework is especially important is dipolar excitons. These excitons are bound pairs of excited electrons ($e^{-}$) and holes ($h^{+}$) generated by photoexcitation. Experimentally, the primary probe of the exciton fluid is its photoluminescence (PL) spectrum, corresponding to the photon energies $E_{\mathrm{ph}}$ emitted during exciton recombination, as schematically illustrated in \cref{fig:illustration}. Because this radiative decay removes a single exciton from the system, the corresponding energy of the emitted photon is directly given by the fluid's chemical potential, $E_{\mathrm{ph}} = \mu_X + \mu(n,T)$ \cite{zimmerman_collective_2022,Rapaport_2020,Fogler_quadrilayer_2021}. Here, the single-exciton part, $\mu_X$, is a reference emission energy determined by the specific setup. The second term, $\mu(n,T)$, is the thermodynamic many-body chemical potential, which is generally density- and temperature-dependent.

One major caveat is that there is no direct experimental probe to measure the exact value of the exciton density. Consequently, the density is typically inferred indirectly via measurements of interaction-driven energy shifts in the PL emission. However, this method faces a major conceptual challenge: accurately extracting the density from observed energy shifts requires prior knowledge of the exciton collective state and spatial correlations, which in turn depend on the unknown density. In practice, to simplify this analysis, most experiments resort to a mean-field approximation of uncorrelated excitons, despite the expectation of strong correlations \cite{zimmermann_excitonexciton_2007,schindler_analysis_2008,lozovik_strong_2007,Remeika2015Measurements,Schinner2013Manybody}. Furthermore, while previous theoretical and experimental studies have primarily focused on the interaction energy \cite{laikhtman_exciton_2009,zimmermann_excitonexciton_2007,Remeika2015Measurements,Schinner2013Manybody,Stern2008Photoluminescence}, a practical calibration scheme must account for the total chemical potential $\mu(n,T)$. Consequently, achieving a reliable and accurate calibration of experimental exciton densities has remained a major unresolved challenge.

In this work, we use large-scale quantum Monte Carlo (QMC) simulations \cite{Buchler2007Strongly,Astrakharchik2007Quantum} to provide a comprehensive and quantitatively controlled test of the analytical framework for two-dimensional dipolar fluids proposed in Ref.~\cite{laikhtman_exciton_2009}, taking dipolar excitons as a prototypical physical realization. We show that the predicted depletion region indeed emerges across the full parameter space, validating the premise of an effectively universal dipolar fluid. Remarkably, the analytical expressions developed in Ref.~\cite{laikhtman_exciton_2009} capture the interaction energy with high accuracy deep within their respective regimes. By contrast, we show that the mean-field approximation deviates significantly from the exact energy, demonstrating its clear invalidity for describing the dipolar system.

We demonstrate that the chemical potential of the dipolar fluid can be accurately modeled across the entire relevant parameter space by a simple, physically intuitive decomposition into three terms: an ideal Bose-gas kinetic contribution that dominates in the dilute, high-temperature limit; a quantum confinement term representing zero-point motion at high densities; and a regime-specific dipolar interaction term. Deep within each correlation regime, this decomposition yields a fully analytical expression for the chemical potential, providing a unified and intuitive description of the strongly correlated dipolar fluid.

Moreover, our findings establish a direct and reliable connection between experimentally measured energy shifts and the underlying density and spatial correlations in the quantum dipolar fluid. To this end, we present practical calibration curves for gallium arsenide (GaAs) and transition-metal dichalcogenide (TMD)-based exciton systems, enabling the accurate determination of density and correlations directly from measured energy shifts. 

The paper is organized as follows: In Section II, we present our model Hamiltonian for the dipolar system and the QMC simulations used to obtain exact values of the two-body correlation function and the dipolar chemical potential. We then present QMC results showing a depletion region in the pair correlation function across a wide parameter space. This observation justifies the universal dipolar description of the system. Building on this dipolar description, in Section III we outline the analytically predicted correlation regimes of the dipolar fluid and their boundaries, and derive the fluid's interaction energy in each regime. We then define an ansatz for the chemical potential, composed of kinetic and interaction contributions, and obtain explicit analytical forms for these terms across the entire phase diagram using the corresponding analytical expressions for the interaction energies. In Section IV, we present numerical results that validate the analytical expressions for both the interaction energy and the chemical potential in each regime, and then use this validation to provide a concrete analytic mapping between energy shifts and exciton densities. Finally, in Section V, we summarize our findings and highlight their relevance for other dipolar systems.

\section{QMC simulation}

We consider a system of $N$ point dipolar bosons in two dimensions, with all dipole moments aligned perpendicular to the plane. The corresponding Hamiltonian reads
\begin{equation}\label{eq:H}
    H = -\frac{\hbar^2}{2m}\sum_{i}\nabla^2_i +\sum_{i> j} U_{dd}\left(|\mathbf{r}_i - \mathbf{r}_j |\right) \,,
\end{equation}
where $\mathbf{r}_i$ is the in-plane position of particle $i$ and $m$ is the particle mass. The first term represents the kinetic energy, and the second term corresponds to the purely repulsive dipolar potential, $ U_{dd}\left(r\right) = D/r^3$.
The dipolar coupling constant, $D$, is defined via $D=e^2 d^2/\kappa$, with $e$ being the electron charge, $d$ the dipole length, and $\kappa$ the dielectric constant. 

To solve \cref{eq:H}, we employ path integral Monte Carlo simulations \cite{Ceperley_1995} using the continuous-space worm algorithm \cite{Boninsegni_2006}. For a given density and temperature, our simulation yields the chemical potential of the dipolar fluid \cite{Del_Maestro_2014}, the two-body correlation function $g_2(r)$, the interaction energy $E_{dd}$, and the superfluid fraction obtained from winding number fluctuations
\cite{Pollock_1987}. We present our results in units of the dipolar length and energy scales $r_0=mD/\hbar^2$ and $E_0 = D/r_0^3$.
Further computational details and observable definitions are provided in  \cref{app:sim}.

We begin by characterizing the spatial correlation profile of the dipolar fluid. To this end, we numerically compute $g_2(r)$ over a wide range of densities, covering the entire experimentally relevant region for excitons, along two representative isotherms: $T_l$ and $T_h$. 
These temperatures are chosen to represent the quantum and classical limits of the fluid, respectively. To define the boundary between them, $T_{\text{Q-C}}$, we consider the effective distance of closest approach, $a_{\text{min}}$, dictated by the repulsive barrier of the dipolar potential.
In the high-temperature limit, this barrier is determined by the distance at which the repulsive potential balances the thermal energy, yielding $a^{\text{C}}_{\text{min}} = (D/k_B T)^{1/3}$. At low temperatures, the particles' wavefunctions overlap at distances larger than this classical barrier, and the effective distance of closest approach is instead dictated by the balance between kinetic energy and dipolar interaction, $a^{\text{Q}}_{\text{min}} = 2mD/\hbar^2=2r_0$. The transition between these regimes occurs when the size of the particle's wavefunction, given by its thermal wavelength $\lambda_T = \sqrt{\hbar^2/2mk_B T}$, exceeds $a^{\text{C}}_{\text{min}}$, yielding $T_{\text{Q-C}} = \hbar^2/8mr_0^2$ \cite{laikhtman_exciton_2009}. Consequently, the probed isotherms are chosen to be $T_l = 0.16 T_{\text{Q-C}}$ and $T_h= 4.9 T_{\text{Q-C}}$. 

\begin{figure*} 
\includegraphics[width=0.93\textwidth]{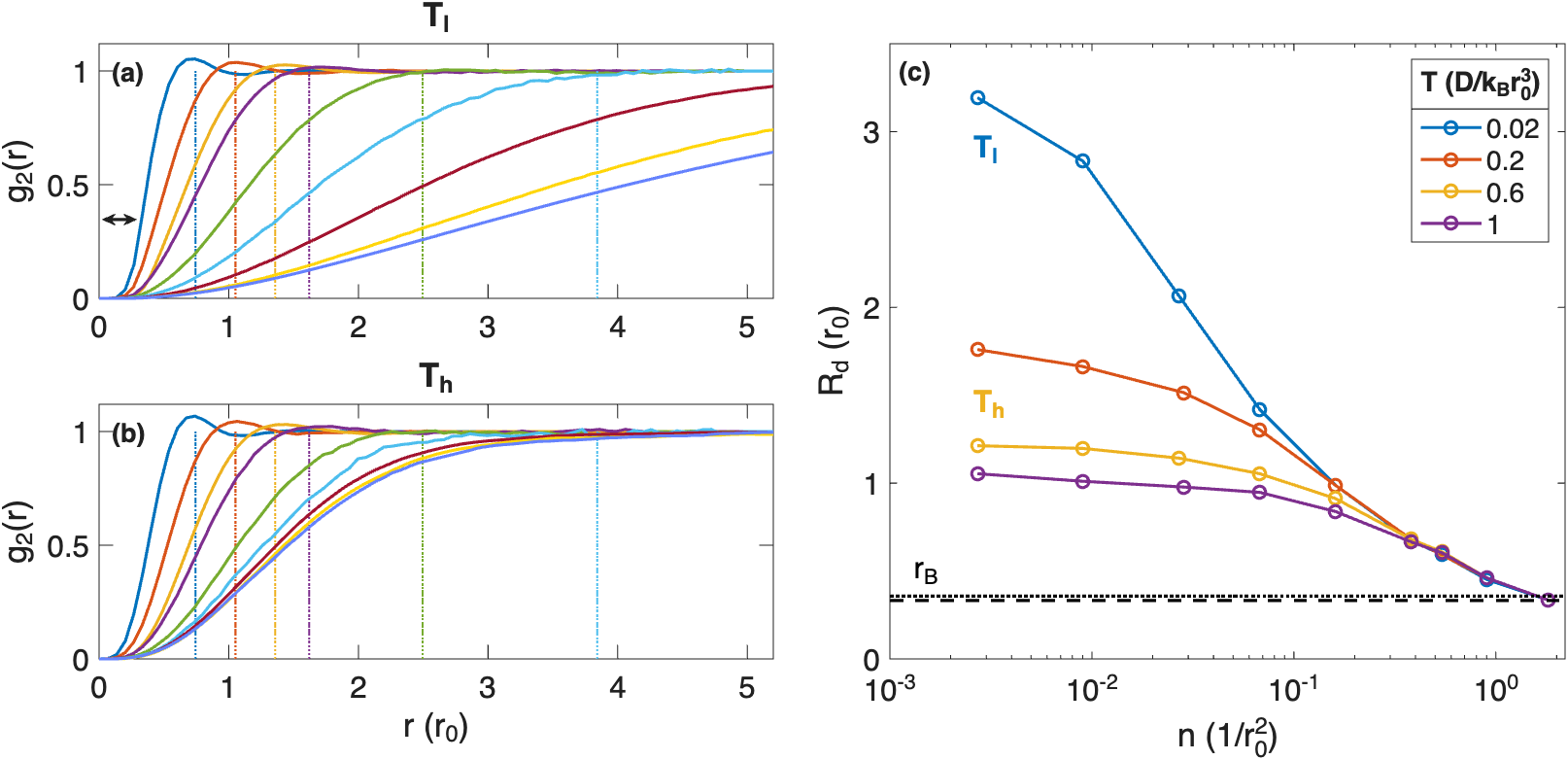}
\centering
\caption{(a, b) Pair correlation functions, $g_2(r)$, obtained via QMC simulations at temperatures $T_l$ and $T_h$, respectively. Solid line colors correspond to the densities sampled along these isothermal trajectories, as indicated by the colored dots in the phase map in \cref{fig:phase_map}. Vertical dotted lines mark the mean interparticle distance, $a = 1/\sqrt{n}$, for each curve. The black arrow highlights the threshold $g_2(r) = 1/e$, from which the depletion radius is extracted. (c) Depletion radius, $R_d$, as a function of density across various temperatures. Horizontal dashed and dotted lines mark the Bohr radius, $r_B$, for GaAs and TMD systems, respectively. 
A typical GaAs system ($d = 10$~nm, $\kappa = 12.9$, $m = 0.21 m_e$) corresponds to characteristic temperatures $T_l = 0.1$~K and $T_h = 3$~K. For a bilayer TMD heterostructure ($d = 1$~nm, $\kappa = 3.3$, $m = 0.5 m_e$), $T_l = 4.8$~K and $T_h = 145$~K. Finally, for comparison, the polar molecule strontium oxide (SrO) (dipole moment of 8.9 Debye, $m \approx 1.89 \times 10^5 m_e$) yields $T_l = 2.7$~K and $T_h = 81$~K \cite{Gadway_2016, Baranov_2012}.  \label{fig:g2_W}}
\end{figure*}

Figs.~\ref{fig:g2_W}a-b show the pair correlation function for several densities, normalized to unity at large separations under the assumption of a homogeneous fluid. We find that $g_{2}(r)$ is strongly depleted at short distances for all simulated densities and temperatures due to strong short-range dipolar repulsion. The apparent correlation hole is the central condition behind the proposed universality: if its radius exceeds the range of exchange, contact, or internal-structure processes, the macroscopic fluid behavior is dictated by the dipolar interaction. For excitons, such suppression quenches short-range exchange and spin conversion, consistent with the stabilization of Bose-condensed long-lived dark populations in earlier experiments \cite{shilo_particle_2013,cohen_dark_2016,Harpaz_dynamical_2019,Misra_2022}. Consequently, this observation justifies modeling excitons as structureless point bosons (\cref{eq:H}). Moreover, the presence of such a significant short-range depletion region invalidates a mean-field description of the dipolar fluid.

The shape of $g_2(r)$ beyond the depletion region distinguishes two correlation regimes. At low density, it rises monotonically toward unity and shows no signatures of extended-range spatial correlations. We identify this behavior as characteristic of a gas phase dominated by two-body scattering processes. We note, however, that $g_2(r)$ exhibits qualitatively different behaviors at the two temperatures. At low temperatures ($T_l$), thermal fluctuations are suppressed, and dipolar repulsion pushes the dipoles farther apart as the fluid becomes more dilute. Consequently, the pair correlation function shifts to larger interparticle separations.  At high temperatures ($T_h$), $g_2(r)$ becomes weakly dependent on density. This originates from the dominance of thermal energy over dipolar interactions\footnote{Moreover, since $T_h > T_{\text{Q-C}}$, this dilute, high-temperature regime is well-described by the classical virial expansion. Within this framework, the pair correlation function is density-independent in the extreme dilute limit \cite{hansen2013theory}, explaining the behavior observed in \cref{fig:g2_W}b.}.

As the density increases, $g_2(r)$ shifts towards smaller distances and, at high enough densities, develops a pronounced peak near the mean interparticle distance, $a=1/\sqrt{n}$. These emergent extended-range correlations mark a crossover into a strongly correlated liquid phase \cite{chandler1987introduction}. 

To quantitatively characterize this crossover, we extract the depletion region radius $R_d$ (schematically depicted in \cref{fig:illustration}), defined as $g_2(R_d)=1/e$, and track its evolution with density and temperature. The results are presented in \cref{fig:g2_W}c. At low densities, $R_d$ decreases with increasing temperature as thermal fluctuations allow dipoles to partially overcome the repulsive barrier. At high densities, different temperature curves converge and $R_d$ becomes temperature-independent. This observation indicates that dipolar interactions strongly suppress thermal fluctuations, providing a clear signature of a strongly correlated dipolar liquid state governed by interparticle correlations. 

The depletion region width defines the validity limit of the dipolar description. To evaluate it for excitonic systems, we compare $R_d$ with the typical Bohr radius, $r_B$, in GaAs and TMD systems, taken as $10$ and $1$~nm, respectively \cite{Sivalertporn_direct_2012,Lobanov_theory_2016,Deng_review_2024}. For typical dipole lengths $d=10$ and $1$~nm and masses $m= 0.21$ and $0.5 m_e$ (with $m_e$ being the free electron mass) in these setups \cite{levinshtein1999handbook,Goryca_2019_revealing,MacDonald2018Theory,Deng_review_2024}, we obtain $r_B/r_0\approx0.36$ and $0.33$, respectively. We find that $R_d$ substantially exceeds $r_B$ at low densities and approaches its value as the density increases. The crossing point $R_d(n)=r_B$ can serve as an estimate of the upper density limit for which our model is valid. At density values exceeding this crossing point, the excitonic wavefunctions overlap, leading to the emergence of short-range physics in the exciton-exciton interactions, and eventually to a Mott transition where excitons dissociate into an electron-hole plasma.

To complete the characterization of the dipoles' collective state, we use our simulations to probe the onset of condensation in the fluid by computing the superfluid stiffness as a function of density and temperature. Our results reveal a transition into a superfluid state at experimentally accessible temperatures and densities, as shown in \cref{fig:phase_map}. For more details on the determination of the critical Berezinskii-Kosterlitz-Thouless (BKT) transition temperature, see \cref{app:data}.

\section{Analytical model}

Building on the QMC results, we now construct an analytical framework for the two-dimensional quantum dipolar fluid. The validity of such a purely dipolar description is grounded in the pronounced depletion region established above. To provide a comprehensive analytical description of the fluid as a function of density and temperature, we first map its distinct correlation regimes, spanning dilute classical and quantum gas phases and a high-density liquid state, and provide explicit expressions for the interaction energy in each state. Next, to connect these predictions to experimental observables, we derive the fluid's chemical potential, $\mu$. To do this, we propose an intuitive ansatz that decomposes $\mu$ into a sum of kinetic and interaction contributions, and show that these components can be derived from the kinetic and interaction energies in each regime. Putting this all together, we show that our analytical ansatz provides a unified description of the fluid across the entire parameter space relevant to exciton experiments.

We begin by evaluating the system's interaction energy. To this end, we consider the interaction energy of a single dipole with the rest of the fluid, which is given by integrating over all pair interactions,
\begin{equation} \label{eq:delta_E_def}
    \epsilon_{dd} = n\int U_{dd}(\mathbf{r}) g_2(\mathbf{r}) d^2\mathbf{r} = 2\pi n \int_0^{L/2} g_2(r) U_{dd}(r) r dr \,.
\end{equation}
Here, $n$ is the density and $U_{dd}(r)$ is the dipolar potential stated in \cref{eq:H}. The second equality follows from the rotational symmetry of $g_2(r)$, with $L$ denoting the system size. In this convention, the total interaction energy of the fluid equals $E_{dd}=N\epsilon_{dd}/2$, correcting for the double-counting of pair interactions, as we verify in \cref{app:data}.

\cref{eq:delta_E_def} explicitly shows the fundamental challenge of experimental density calibration. Because exciton densities are typically extracted from the interaction-driven blueshift, accurate density calibration requires prior knowledge of the fluid's spatial correlations, represented by $g_2(r)$. However, as these correlations are highly sensitive to density, temperature, and dipole length, solving this integral requires mapping the different correlation regimes of the fluid.

The simplest assumption employs a mean-field (MF) treatment, in which interparticle correlations are completely neglected and $g_2(r)=1$. This approach yields the so-called "plate capacitor formula"\footnote{When $g_2(r)=1$, the point-dipole approximation yields a diverging energy integral due to short-distance particle overlap. To regularize this integral, the particles must be treated as extended dipoles using the Coulomb interaction $U_{dd}(r)=\frac{2e^2}{\kappa}\left(\frac{1}{r} - \frac{1}{\sqrt{r^2 + d^2}}\right)$.},
\begin{equation} \label{eq:MF_model}
    \epsilon^{\text{MF}}_{dd} = \frac{4\pi e^2 d n}{\kappa} \,.
\end{equation}
Although widely used in earlier exciton studies \cite{Remeika2015Measurements,Sun2022Excitonic,Schinner2013Manybody,Remeika_2009_localization,Alloing_2014_evidence,Anankine_2017}, this uncorrelated assumption is invalid for the dipolar fluid, as dictated by the short-range depletion of $g_2(r)$ (see \cref{fig:g2_W}).

\begin{figure} 
\includegraphics[width=0.49\textwidth]{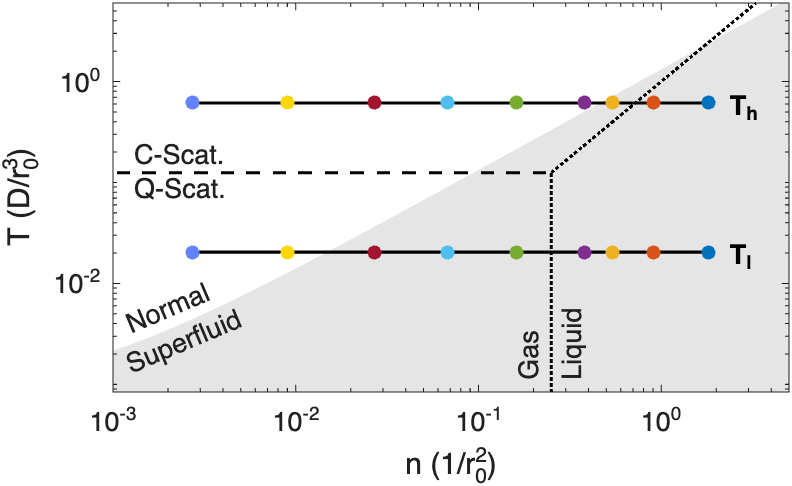}
\centering
\caption{(a) The phase diagram showing different correlation regimes in the dipolar fluid as a function of density and temperature. The dotted black line marks the predicted transition between two-particle correlations (gas phase) and multiparticle correlations (liquid phase). The dashed black line marks the predicted transition between classical and quantum scattering in the gas phase. QMC calculations are performed along the two isothermal trajectories of $T_l$ and $T_h$. The colored dots mark QMC-sampled densities, corresponding to the coloring scheme used for the $g_2(r)$ curves in \cref{fig:g2_W}a-b. The shaded area marks the superfluid regime extracted from QMC calculations. \label{fig:phase_map}}
\end{figure}

Following this line of reasoning, Laikhtman and Rapaport~\cite{laikhtman_correlations_2009,laikhtman_exciton_2009} developed a general theoretical framework that predicts distinct correlation regimes in the dipolar fluid, based on the fundamental energy and length scales of the system. The resulting phase map, drawn for realistic densities and temperatures of excitonic fluids, distinguishes between liquid and gas characteristics as a function of density. Moreover, it identifies two scattering regimes within the gas phase as a function of temperature. At high temperatures, quantum effects are suppressed, and the gas is dominated by classical two-body scattering processes (C-Scat.), while at low temperatures the scattering becomes of a quantum nature (Q-Scat.). These regimes, marked on the density-temperature phase map in \cref{fig:phase_map}, show distinct correlation patterns that determine the corresponding expressions for the fluid's interaction energy.

In the classical gas regime (C-Scat.), the pair correlation function can be described via the virial expansion \cite{Pathria_statmech_2011}. In the $n\to0$ limit, three-body collisions vanish, and spatial correlations are governed by the pair potential and the temperature, $g_2(r) = e^{-U_{dd}(r)/k_BT}$. Although finite densities introduce higher-order virial terms that render $g_2(r)$ density-dependent, this effect remains negligible deep within the gas phase, as shown numerically in \cref{fig:g2_W}b. Supported by this observation, we substitute the above lowest-order form of $g_2(r)$ into the energy integral to obtain the interaction energy of the classical gas \cite{laikhtman_exciton_2009},
\begin{equation} \label{eq:CDG_model}
    \epsilon^{\text{CG}}_{dd} = 2\pi \Gamma(4/3) D^{2/3} (k_BT)^{1/3} n \,.
\end{equation}

As temperature decreases, quantum effects in particle scattering become dominant as the thermal wavelength $\lambda_T$ exceeds $a^{\text{C}}_{\text{min}}$. This occurs at temperatures below $T_{\text{Q-C}}$. Quantum treatment of the low-energy dipolar scattering problem yields the following interaction energy for small scattering wavenumbers \cite{laikhtman_exciton_2009},
\begin{equation} \label{eq:QDG_model}
    \epsilon^{\text{QG}}_{dd} = \frac{2\pi \hbar^2}{m \ln^2(k_T r_0)} n\,.
\end{equation}
where $k_T = 1/\lambda_T$ is the thermal wave vector.

With the transition to the high-density limit, multiparticle scattering becomes substantial, leading to extended-range spatial correlations in the fluid. This scenario is manifested by the emergence of a nearest-neighbor peak in $g_2(r)$ (see Figs.~\ref{fig:g2_W}a-b), characteristic of a liquid. To evaluate the threshold density of the gas-to-liquid transition, we compare the mean interparticle distance, $a$, to the size of the effective distance of closest approach, $a_{\text{min}}$. At high temperatures, this barrier is given by $a^{\text{C}}_{\text{min}}$, and the fluid transitions to a liquid when the density exceeds $(a^{\text{C}}_{\text{min}})^{-2}$, giving a gas-liquid boundary at $n_{\text{CG-L}} = (k_B T/D)^{2/3}$ \cite{laikhtman_exciton_2009}. Conversely, at low temperatures, the closest approach is dictated by $a^{\text{Q}}_{\text{min}}$, corresponding to a liquid transition at $n_{\text{QG-L}} = (a^{\text{Q}}_{\text{min}})^{-2}=1/(4r_0^2)$ \cite{laikhtman_exciton_2009}. 

To quantify the fluid's interaction energy in the above high-density regime, we assume that each dipole is confined within an effective potential well generated by the strong repulsion of the surrounding fluid. In this case, the lowest-order contribution to the interaction energy comes from dipolar interactions between localized dipoles at the bottom of the wells, exhibiting a short-range crystalline order. For a triangular lattice ordering, this static component was shown to be \cite{laikhtman_exciton_2009}
\begin{equation} \label{eq:static_model}
    \epsilon_{\text{static}} = 10Dn^{3/2} \,.
\end{equation}
Beyond the static limit, the second-order contribution to the interaction energy accounts for spatial fluctuations of the dipoles within effective harmonic potentials formed by the repulsive environment, characterized by a zero-point positional fluctuation scale $R_Q$ (see \cref{fig:illustration}). This term yields zero-point quantum fluctuations at the bottom of the harmonic wells, described by \cite{laikhtman_exciton_2009}
\begin{equation} \label{eq:EQ_model}
    \epsilon_Q = \sqrt{\frac{30D^2}{r_0}} n^{5/4} \,.
\end{equation}
Since $\epsilon_Q$ corresponds to the quantum ground-state energy of a two-dimensional harmonic oscillator, the virial theorem dictates that it consists of equal potential and kinetic energy components, $V_Q = T_Q = \epsilon_Q/2$. The kinetic term $T_Q$ corresponds to the zero-point motion energy of a single dipole in the dense limit of the liquid. The potential term $V_Q$ represents the energy penalty of the dipole's motion against its repulsive cage and should be accounted for to properly describe the interaction energy of the liquid. Combining the static and harmonic contributions yields the following interaction energy for the liquid state,
\begin{equation} \label{eq:liquid_model}
    \epsilon^{\text{L}}_{dd} = \epsilon_{\text{static}} + V_Q = \epsilon_{\text{static}}+\epsilon_{\text{Q}}/2 \,.
\end{equation}

Having established the distinct correlation regimes of the dipolar fluid and their corresponding interaction energies, we now connect these results to experimental observables. In excitonic systems, the primary experimental probe is the interaction-induced energy shifts. Because an emitted photon carries the energy of a single recombining exciton leaving the interacting system, the spectral shifts correspond directly to the exciton chemical potential, with the emission energy given by $E_{\mathrm{ph}} = \mu_X + \mu(n,T)$. Focusing on the many-body term, $\mu$, it is formally defined as the derivative of the total free energy with respect to particle number, $\mu = \partial F/\partial N$. However, this form generally lacks closed analytical expressions for strongly correlated quantum systems. To provide a tractable analytical model, we propose an intuitive ansatz that decomposes the chemical potential into a sum of kinetic ($\mu_k$) and interaction ($\mu_{dd}$) terms:
\begin{equation} \label{eq:mu_ansatz}
\mu = \mu_{k} + \mu_{dd} \,,
\end{equation}
The specific forms of these components naturally depend on the fluid's state. Below, we derive them for each correlation regime.

We begin with the dilute gas limit. In the absence of many-body correlations, the kinetic component is governed by the kinetic energy of free particles, taking the form of an ideal two-dimensional Bose gas, $\mu_k = \mu_{BE}$, where
\begin{equation} \label{eq:mu_BE}
    \mu_{BE} = k_BT \ln\left[1 - \exp\left(-\frac{2\pi\hbar^2 n}{mk_BT}\right)\right] \,.
\end{equation}

To evaluate the corresponding interaction term, we first consider the classical gas regime. In this limit of high temperature and low density, $\mu_{dd}$ can be directly derived from the classical virial expansion \cite{Pathria_statmech_2011}. Carrying out this calculation (see \cref{app:data}) yields a proportionality relation between $\mu_{dd}$ and the interaction energy of the classical gas, where $\mu_{dd} = \alpha \epsilon^{\text{CG}}_{dd}$, with $\alpha = 3/2$ \cite{laikhtman_exciton_2009}. Thus, in the classical gas regime, the total chemical potential is explicitly given by $\mu = \mu_{BE} + \frac{3}{2}\epsilon^{\text{CG}}_{dd}$.

As the temperature decreases, the fluid enters the quantum gas regime. Because thermal fluctuations are strongly suppressed in this limit, the system's free energy is governed by its internal energy. Thus, in this limit, $\mu_{dd}$ can be approximated as the particle-number derivative of the interaction energy, i.e., $\mu_{dd} = \frac{\partial}{\partial N}\left(E_{dd}\right)$, where $E_{dd}=\frac{N}{2}\epsilon_{dd}$ and $\epsilon_{dd}$ takes the quantum gas form of \cref{eq:QDG_model} ($\epsilon_{dd}=\epsilon^{\text{QG}}_{dd}$).
Since the interaction energy scales as a power law with density, $\epsilon^{\text{QG}}_{dd} \sim n^{\gamma}$, the internal-energy derivative yields a proportionality relation of the form $\mu_{dd} = \alpha \epsilon^{\text{QG}}_{dd}$, where now the scaling factor is given by $\alpha = \frac{1}{2}(\gamma+1)$. The linear dependence of \cref{eq:QDG_model} on the density ($\gamma=1$) yields $\alpha = 1$. Consequently, the chemical potential for the quantum gas takes the form of $\mu = \mu_{BE} + \epsilon^{\text{QG}}_{dd}$.

In the high-density limit, the fluid transitions into the strongly interacting liquid state. To evaluate $\mu_{dd}$ in this regime, we note that the fluid's interaction energy vastly exceeds its thermal energy, placing the system in an effective zero-temperature limit (see \cref{app:data}). Thus, the internal-energy derivative approximation for $\mu$, introduced for the quantum gas, can be applied for the liquid as well. In this case, the interaction energy (\cref{eq:liquid_model}) comprises a static lattice contribution, $\epsilon_{\text{static}} \propto n^{3/2}$, and a potential contribution, $V_Q \propto n^{5/4}$. Applying the scaling relation $\alpha = \frac{1}{2}(\gamma+1)$ to these components yields $\alpha = 5/4$ for the static term and $\alpha = 9/8$ for the potential term. Therefore, the interaction chemical potential in the liquid state is $\mu_{dd} = \frac{5}{4}\epsilon_{\text{static}} + \frac{9}{8}V_Q$.

Concurrently, the kinetic component of the chemical potential, $\mu_k$, can no longer be described via the ideal gas expression as the spatial overlap of the particles' wavefunctions becomes substantial. Instead, the effective confinement of the dipoles, formed by strong dipolar interparticle repulsion, gives rise to kinetic zero-point fluctuations, with kinetic energy $T_Q$, at the bottom of the confining potentials. Because the effective confinement cage of each dipole is formed by the rest of the particles in the fluid, the total contribution of this effect to the fluid's kinetic energy is $NT_Q/2$, with the $1/2$ factor preventing double counting. Employing the internal-energy derivative method to this term, we obtain an analytical expression for the high-density kinetic component of the chemical potential, $\mu_k = \mu_Q = \frac{\partial}{\partial N}\left(\frac{N}{2}T_Q\right)$. Since $T_Q$ scales as $n^{5/4}$, we get $\mu_Q=\frac{9}{8}T_Q=\frac{9}{16}\epsilon_{\text{Q}}$.
Bringing this all together, the total chemical potential of the liquid is $\mu = \mu_Q + \mu_{dd} = \mu_Q + \frac{5}{4}\epsilon_{\text{static}} + \frac{9}{8}V_Q $.

To construct a unified analytical description of the quantum dipolar fluid, we note that the two kinetic terms, $\mu_{BE}$ and $\mu_Q$, provide complementary limits for the kinetic contribution to the chemical potential. Since the ideal-gas contribution $\mu_{BE}$ vanishes at high densities and the zero-point quantum fluctuation term $\mu_Q$ vanishes at low densities, they can be combined to define the kinetic component as $\mu_k = \mu_{BE} + \mu_Q$. In this way, $\mu_k$ is valid across the low- and high-density regimes. Substituting this into our initial ansatz (\cref{eq:mu_ansatz}) yields the final, comprehensive equation for the chemical potential:
\begin{equation} \label{eq:mu_decomp}
\mu = \mu_{BE} + \mu_{dd} + \mu_Q \,,
\end{equation}
where the specific form of $\mu_{dd}$ is dictated by the corresponding correlation state of the fluid. Thus, this single equation provides a unified description of the quantum dipolar fluid across a wide range of densities and temperatures.

With this theoretical framework established, we now turn to its quantitative validation. In the following section, we utilize our exact QMC tool to systematically verify both the regime-specific predictions for the interaction energy and the proposed decomposition of the chemical potential.

\section{Results}

\subsection{Quantitative Verification of the Analytical Energy Models}

\begin{figure*}
\includegraphics[width=0.98\textwidth]{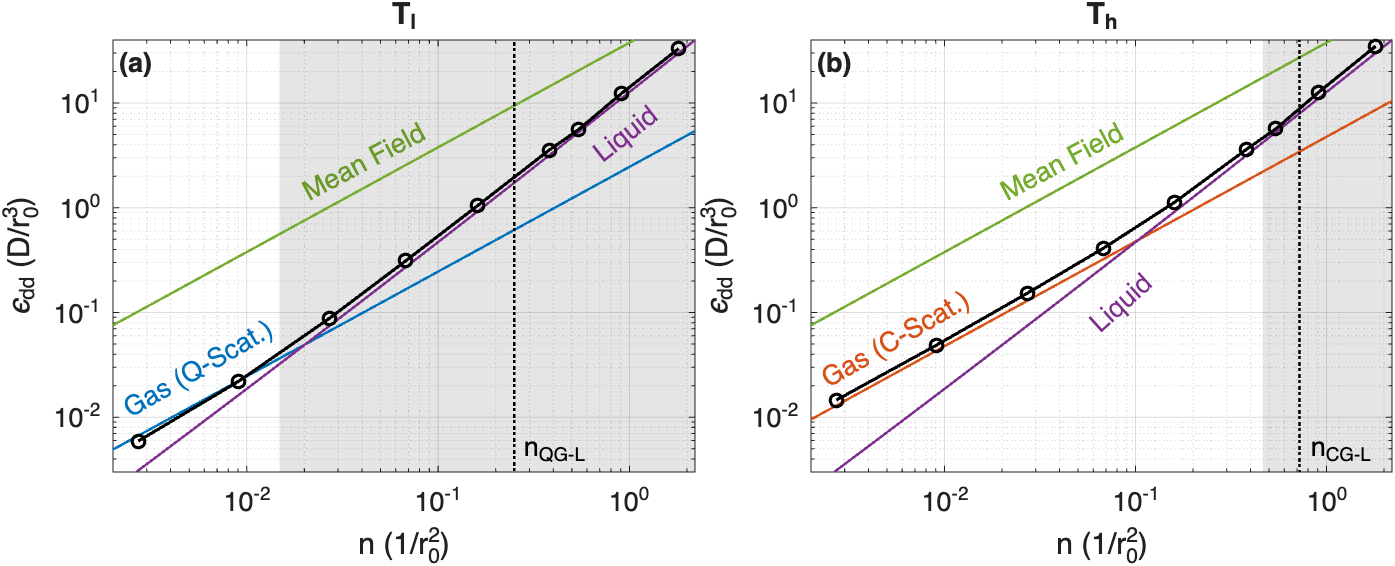}
\centering
\caption{The QMC-calculated dipolar interaction energy $\epsilon_{dd}$ (black circles) compared with analytical models (solid lines) in distinct correlation regimes, as shown for (a) $T_l$ and (b) $T_h$. At low densities, $\epsilon_{dd}$ is well-described by the asymptotic gas models, governed by quantum scattering at $T_l$ (Q-Scat., blue) and classical scattering at $T_h$ (C-Scat., red). Vertical dashed lines mark the theoretical gas-to-liquid transition densities, $n_{\text{QG-L}}$ at $T_l$ and $n_{\text{CG-L}}$ at $T_h$. The shaded area indicates the superfluid regime. At high densities, $\epsilon_{dd}$ converges to the liquid model (purple) at both temperatures. Across all relevant densities, the mean-field approximation (green) significantly overestimates $\epsilon_{dd}$ compared to the exact QMC results. \vspace{0.2cm}} \label{fig:dE}
\end{figure*}

To quantitatively verify the above ansatz for $\mu$, we first test the analytical expressions for the interaction energy of the dipolar fluid: we compute $\epsilon_{dd}$ from \cref{eq:delta_E_def} using the exact QMC data for $g_2(r)$, presented in \cref{fig:g2_W}.
Then, in \cref{fig:dE}, we compare these numerical results with the analytical expressions for $\epsilon_{dd}$ as a function of density along both the $T_l$ and $T_h$ isothermal trajectories. We find that the analytical models provide excellent approximations within their respective regimes, accurately capturing the distinct power-law scalings of the energy. At low densities, the numerical $\epsilon_{dd}$ values tightly follow the linear density scaling ($\propto n$) of the gas models, with quantum scattering at $T_l$ and classical scattering at $T_h$. As the density increases, the QMC data systematically shift from the gas models toward the liquid model, characterized by leading $\propto n^{3/2}$ static scaling and $\propto n^{5/4}$ zero-point corrections. The error in the analytical models peaks near the crossing points of the gas and liquid expressions, reaching about 17\% at $T_l$ and 30\% at $T_h$. However, the errors are small deep within the regimes where each model applies, typically falling below 10\%. 
Consequently, the theoretically predicted phase boundaries, $n_{\text{QG-L}}$ and $n_{\text{CG-L}}$ (marked by dashed lines in \cref{fig:dE}), remain solid theoretical reference points for determining gas and liquid behaviors. Although they may overestimate the crossover onset by up to an order of magnitude, they are derived from the underlying characteristic length scales of the fluid. As such, they provide a simple and intuitive anchor for experimentalists. We note that the numerical QMC data in \cref{fig:dE} indicate that the gas-liquid crossover occurs more accurately near the intersection points $\epsilon_{dd}^{\text{QG}}=\epsilon_{dd}^{\text{L}}$, $\epsilon_{dd}^{\text{CG}}=\epsilon_{dd}^{\text{L}}$, respectively. 

Finally, comparing these exact numerical results with the mean-field approximation (\cref{eq:MF_model}) reveals the severe limitations of the uncorrelated assumption. By assuming a uniform spatial distribution that allows overlapping particle configurations, the mean-field approach ignores the pronounced short-range depletion region in $g_2(r)$. As a result, this model overestimates the interaction energy by more than an order of magnitude at low densities and by a factor of three at the highest realistic exciton densities. This large quantitative discrepancy shows that accurately accounting for many-body correlations across all experimental regimes is necessary.

\subsection{Validation of the Chemical Potential Decomposition}
\vspace{0.2cm}

Having verified the validity of the analytical interaction energies within their respective regimes, we now turn to test the chemical potential decomposition proposed in \cref{eq:mu_decomp}. Here, we compute the individual contributions to $\mu$ using their previously established derivations. While the expressions for $\mu_{BE}$ and $\mu_Q$ remain the same across the parameter regime, the interaction term $\mu_{dd}$ is regime-dependent. The choice of which $\mu_{dd}$ form to use is set by the analytical boundaries $T_{\text{Q-C}}$, $n_{\text{QG-L}}$, and $n_{\text{CG-L}}$.

Since the transition between the gas and liquid regimes is a continuous crossover, the asymptotic analytical models naturally lose accuracy near these theoretical boundaries. To account for this, we define a crossover window around the density boundaries ($n_{\text{QG/CG-L}}$), chosen to span from $n_{\text{QG/CG-L}}/3$ to $1.5n_{\text{QG/CG-L}}$. Outside this window, we apply the asymptotic analytical expressions of the relevant correlation models.
Specifically, for $T<T_{\text{Q-C}}$, we obtain $\mu_{dd}$ using the quantum gas model (\cref{eq:QDG_model}) at densities $n<n_{\text{QG-L}}/3$ and the liquid model (\cref{eq:liquid_model}) when $n>1.5n_{\text{QG-L}}$. Conversely, for $T>T_{\text{Q-C}}$ we compute $\mu_{dd}$ via the energy of a classical gas (\cref{eq:CDG_model}) at $n<n_{\text{CG-L}}/3$, or the energy of a liquid at $n>1.5n_{\text{CG-L}}$.

Within the crossover window, both gas and liquid behaviors are expected to contribute to the fluid's chemical potential. Thus, to ensure a smooth transition of $\mu_{dd}$ across the gas-liquid boundary, we interpolate it between the gas and liquid models. Since the fluid's density spans multiple orders of magnitude, we perform this interpolation linearly on a logarithmic density scale, computing $\mu_{dd}$ via
\begin{equation}
    \mu_{dd}(n) = (1 - w(n)) \mu_{dd}^{\text{QG/CG}}(n) + w(n) \mu_{dd}^{\text{L}}(n) \,,
\end{equation}
where the interpolation weight is defined as
\begin{equation}
    w(n)=\frac{\log(n)-\log(n_1)}{\log(n_2)-\log(n_1)} \,,
\end{equation}
with $n_1$ and $n_2$ being the lower and upper bounds of the interpolation window, respectively. Summing the three terms in the ansatz, we compare the resulting analytical chemical potential with the exact QMC result for $\mu$.

\begin{figure*}
\includegraphics[width=0.95\textwidth]{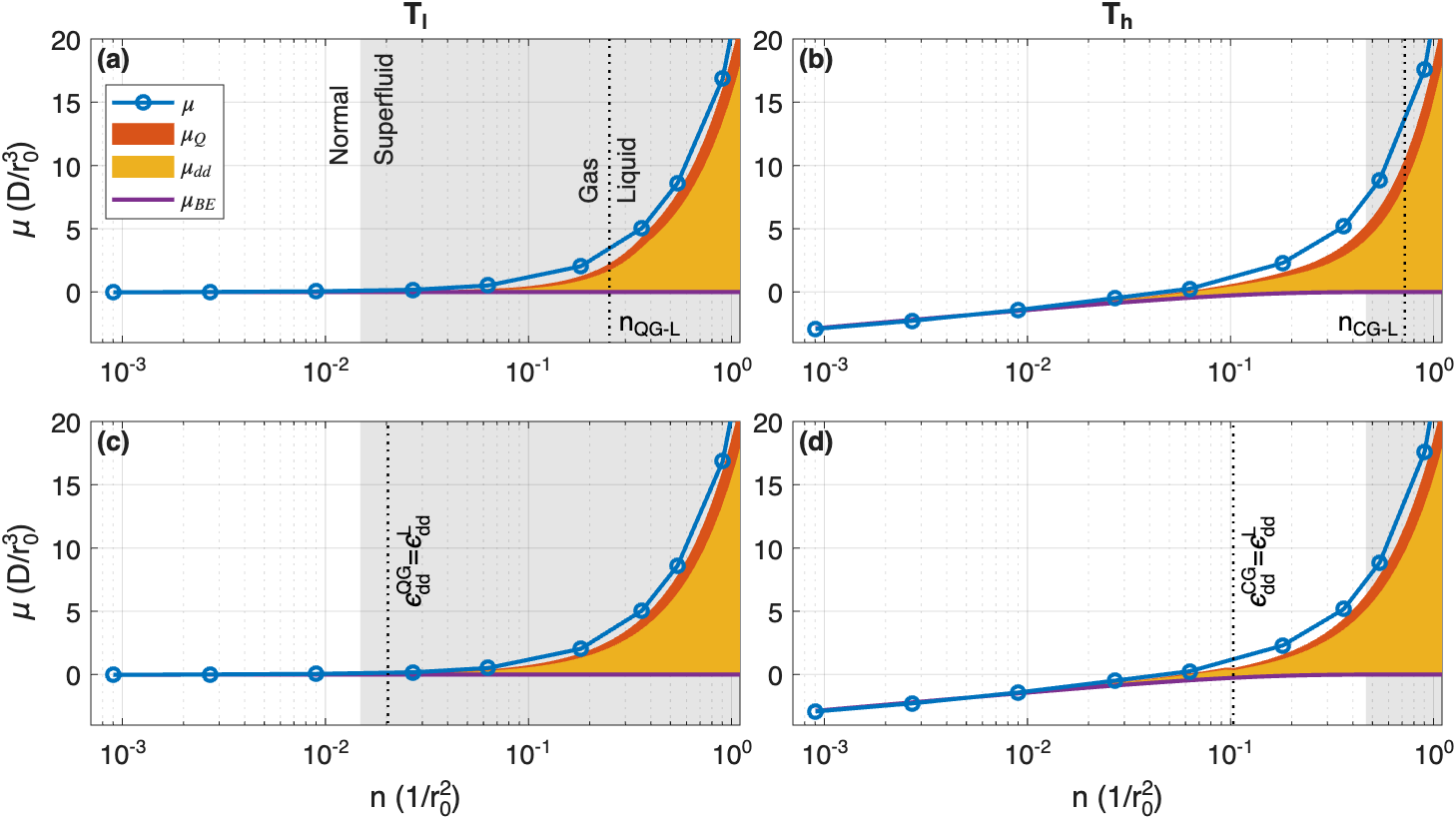}
\centering
\caption{The chemical potential $\mu$ of the dipolar fluid as a function of density at (a, c) $T_l$ and (b, d) $T_h$. Exact QMC results (blue dots, computed for $N=32$ particles) are compared against the analytical reconstruction. The analytical chemical potential is the sum of the ideal Bose gas contribution ($\mu_{BE}$, purple line), the many-body interaction term ($\mu_{dd}$, yellow area), and the zero-point quantum confinement term ($\mu_Q$, red area). The shaded background indicates the superfluid regime.
In panels (a) and (b), vertical dotted lines mark the theoretical gas-liquid phase boundaries, $n_{\text{QG-L}}$ and $n_{\text{CG-L}}$. To model $\mu_{dd}$ in these panels, we use the quantum (a) and classical (b) gas expressions for $n < n_{\text{QG/CG-L}}/3$, and the liquid expression for $n > 1.5n_{\text{QG/CG-L}}$. In the intermediate crossover region ($n_{\text{QG/CG-L}}/3 \le n \le 1.5n_{\text{QG/CG-L}}$), $\mu_{dd}$ is computed via linear interpolation (on the logarithmic scale) between the respective gas and liquid models.  In panels (c) and (d), the transition between the gas and liquid regimes is set by the crossing point of their respective energies, $\epsilon^{\text{QG/CG}}_{dd}=\epsilon_{dd}^{\text{L}}$ (black dotted lines), where the quantum (c) and classical (d) gas expressions are used, respectively. No interpolation is applied in these panels.\vspace{0.2cm}}  \label{fig:mu}
\end{figure*}

As shown in Figs.~\ref{fig:mu}a-b, the analytical decomposition accurately follows the numerical data as a function of density along both isotherms. At low densities, interparticle interactions are negligible, and the chemical potential is captured by the ideal gas term, $\mu_{BE}$. This behavior is most pronounced at higher temperatures, where $\mu$ attains the negative values expected for a two-dimensional ideal Bose gas. In contrast, at low temperatures, $\mu \to 0$, mirroring the asymptotically vanishing trend of $\mu_{BE}$ in this regime.

As density increases, dipole-dipole interactions, expressed by $\mu_{dd}$, become the dominant contribution to the total chemical potential at both temperature limits. At the same time, quantum confinement effects, represented by $\mu_Q$, provide an additional, non-negligible contribution, reflecting the strongly correlated nature of the dipolar fluid in the high-density regime.

Consistent with our previous observations of the interaction energy (\cref{fig:dE}), the analytical model of $\mu$ agrees well with the numerical data deep within the distinct gas and liquid states. However, notable deviations appear at intermediate densities. These deviations stem from using the theoretical phase boundaries, $n_{\text{QG-L}}$ and $n_{\text{CG-L}}$, which overestimate the crossover onsets, to dictate the appropriate functional form of $\mu_{dd}$. To improve the fit of the analytical model to the QMC results in this region, one could instead set the transition density at the intersection of the gas and liquid energy curves. 
Using these energy crossing points as sharp boundaries for $\mu_{dd}$, applying no intermediate interpolation between the models, yields a slightly tighter fit to the QMC data in the crossover region, as shown in Figs.~\ref{fig:mu}c-d.

\subsection{Practical Calibration Curves}

\begin{figure*}
\includegraphics[width=0.9\textwidth]{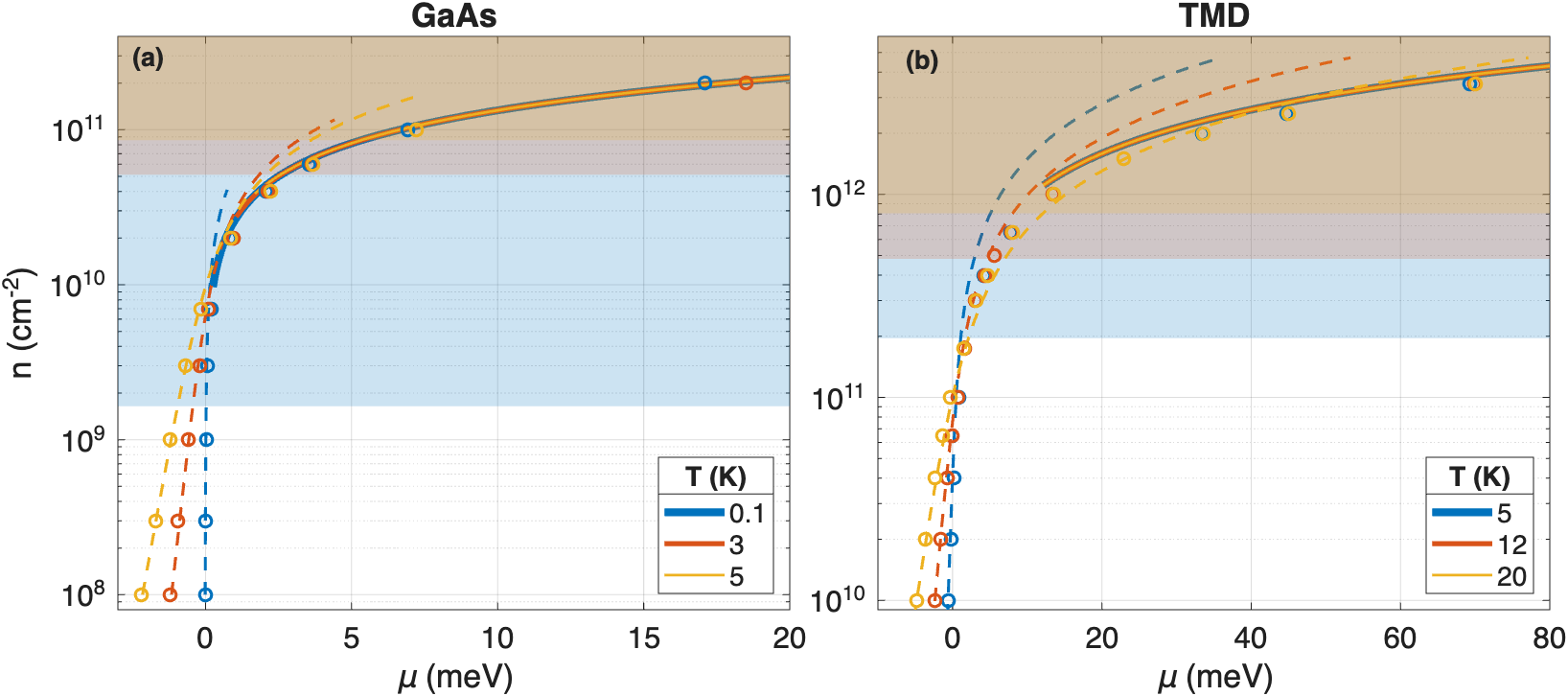}
\centering
\caption{Density calibration curves for common material properties of (a) GaAs (d=10~nm, $\kappa$=12.9, $m$=0.21$m_0$) and of (b) TMD (d=1~nm, $\kappa$=3.3, $m$=0.5$m_0$). Circles mark the exact QMC results for $N=32$. Dashed and solid lines represent the analytical gas and liquid models, respectively. Solid lines are plotted with different widths to make their overlap easier to visualize. The shaded areas mark the superfluid regions at each respective temperature. For the gas phase, the model transitions from the quantum to the classical expression at the threshold temperature $T_{\text{Q-C}}$. To highlight the gas-liquid crossover region, both analytical models are plotted simultaneously across the estimated transition boundary $n_{\text{QG/CG-L}}$, spanning the range $n_{\text{QG/CG-L}}/3 \leq n \leq 1.5n_{\text{QG/CG-L}}$. Outside this crossover window, only the valid model is shown.\vspace{0.1cm}} \label{fig:calib}
\end{figure*}

The above results have immediate practical implications for experimental studies of dipolar exciton fluids. This validated theoretical framework allows for the direct determination of the exciton density and \textit{correlation state} from experimentally measured energy shifts, using the analytical equation for the chemical potential (\cref{eq:mu_decomp}) and the regime-dependent energy expressions (\cref{eq:CDG_model,eq:QDG_model,eq:liquid_model}).

To guide current experiments and address the density calibration challenge, we map exciton density as a function of chemical potential at several temperatures, using typical material parameters for GaAs and TMD heterostructures (\cref{fig:calib}). For a given calibration isotherm, the appropriate low-density gas model is set by comparing the experimentally tuned temperature to the $T_{\text{Q-C}}$ boundary. However, since the density is the unknown target variable, determining whether the fluid resides in the gas or liquid regime cannot be done a priori.

To resolve this, we plot both the corresponding gas and liquid models for each temperature within their expected regimes of validity. Moreover, since the intermediate-density regime is not well-described by the asymptotic models, we extend both curves beyond the theoretical phase boundary ($n_{\text{QG-L}}$ or $n_{\text{CG-L}}$), spanning the range of $n_{\text{QG/CG-L}}/3 \leq n \leq 1.5n_{\text{QG/CG-L}}$. These two overlapping analytical lines define a practical uncertainty window for density calibration in the intermediate-density regime. Beyond this region, the asymptotic gas and liquid equations reliably describe the fluid's chemical potential.

To benchmark the gas-liquid crossover at each temperature, we include the QMC results for $\mu$, which demonstrate excellent agreement with the analytical lines. Consequently, these calibration maps give experimentalists a straightforward, accurate tool to determine exciton densities directly from optical emission spectra, without relying on complex numerical models. 

Notably, our analysis also provides an insightful guideline on the correlation state of the exciton systems: from \cref{fig:calib} we can see that for GaAs-based dipolar excitons at $T=5$K, the fluid is already liquid-like for densities larger than $\simeq10^{10}$~cm$^{-2}$, while for typical TMD-based systems, multiparticle correlations emerge at much higher densities, around $n\simeq2-3\times10^{12}$~cm$^{-2}$. 

Finally, these maps highlight the different temperature scales of the two platforms. TMD systems are characterized by a higher $T_{\text{Q-C}}$, and thus remain in the quantum regime even up to $T=20$K. This is visually evident in \cref{fig:calib}b, where the liquid onset is governed by the temperature-independent quantum boundary, $n_{\text{QG-L}}$ (causing the solid lines to overlap). In contrast, GaAs experiments at $5-3$K operate above their $T_{\text{Q-C}}$, placing them in the classical regime where the liquid onset ($n_{\text{CG-L}}$) shifts with temperature, and enter the quantum regime only as the temperature is lowered to $T=0.1$K.

\section{Conclusion}

In this work, we have established a comprehensive and quantitatively validated analytical framework for two-dimensional quantum dipolar fluids. By mapping the distinct correlation regimes of the fluid across its density-temperature phase diagram, we derived explicit analytical expressions for the system's interaction energy in the classical gas, quantum gas, and liquid regimes. Building on this, we demonstrated that the chemical potential can be accurately described across the entire parameter space via an intuitive three-component decomposition: an ideal Bose gas contribution, a many-body interaction term, and a high-density quantum confinement term. Using exact QMC simulations, we rigorously validated this theoretical model, showing excellent agreement with numerical data deep in the asymptotic gas and liquid regimes.

Crucially, our numerical results confirm the existence of a pronounced short-range depletion region in the two-body correlation function throughout the studied parameter space. This depletion effectively quenches short-range processes, supporting the idea of emergent universality in the fluid, where its collective state is governed by long-range dipolar interactions. Furthermore, this finding exposes the severe limitations of treating the dipolar fluid at the mean-field level. By ignoring spatial correlations, the mean-field approach overestimates interaction energies by up to an order of magnitude, underscoring the need to properly account for these correlations.

By applying our analytical framework to dipolar excitons, we resolved the long-standing experimental challenge of density calibration. Specifically, our framework directly links measured PL energy shifts to exciton densities across all correlation regimes. Leveraging this result, we provided practical calibration maps for representative GaAs and TMD heterostructures to guide ongoing experiments in this field.

The unified analytical description presented here provides both a fundamental understanding of strongly correlated dipolar phases and a reliable tool for resolving immediate experimental challenges. Looking forward, because our model is formulated for point dipolar bosons, these results are widely applicable to other two-dimensional platforms beyond excitonic systems, such as ultracold polar molecules and magnetic atoms, paving the way for broad experimental exploration across diverse quantum systems.

\begin{acknowledgments}
   R.R. acknowledges support from the Israel Science Foundation (ISF) Grant no. 1087/22. S.G. acknowledges support from the Israel Science Foundation (ISF) Grant no. 586/22. M.Z. acknowledges the support of the Council for Higher Education Scholarship Program for Outstanding Doctoral Students in Quantum Science and Technology.
\end{acknowledgments}

\onecolumngrid
\appendix
\section{QMC simulation \label{app:sim}}

In this section, we detail the numerical estimators used in the QMC simulations to extract the physical quantities discussed in the main text. The simulations are performed using the continuous-space worm algorithm \cite{Boninsegni_2006} in the path-integral representation \cite{Ceperley_1995}. These methods allow direct sampling of correlation functions, interaction energy, chemical potential, and superfluid stiffness \cite{Ceperley_1995,Del_Maestro_2014}.

To characterize the spatial correlations in the fluid, we compute the pair correlation function, defined via
\begin{equation}
     g_2(\mathbf r)=\frac{L^2}{N^2}
     \left\langle\sum_{i\ne j}\delta^{(2)}(\mathbf r-\mathbf r_i+\mathbf r_j)\right\rangle \,,
\end{equation}
where $L$ is the linear simulation box size. Using the rotational symmetry of the dipolar Hamiltonian in \cref{eq:H}, we average over the angular coordinate to obtain the radial correlation function $g_2(r)$ shown in the main text. Additionally, assuming a homogeneous fluid, we normalize $g_2(r)$ such that $g_2(r\to \infty) \to 1$.

The total interaction energy of the fluid can be computed using our simulations by averaging over all pair interactions in a given path integral configuration, 
\begin{equation}
    E_{dd}=\left\langle \frac{1}{2} \sum_{i\neq j} U_{dd}(|\mathbf{r}_i - \mathbf{r}_j|)\right\rangle \,,
\end{equation}
where $U_{dd}(r)=D/r^3$ is the dipolar potential and the factor of $1/2$ prevents double counting. 
This estimator measures the total interaction energy of the fluid and serves as an exact numerical benchmark for the analytical integral approach in \cref{eq:delta_E_def}, as discussed in the following section.

To benchmark our analytical ansatz for the fluid's chemical potential, stated in \cref{eq:mu_decomp}, we sample $\mu$ directly from the QMC simulation using the methodology developed in Ref.~\cite{Del_Maestro_2014}. In this way, the chemical potential is extracted from the relative statistical weights of path-integral sectors with different particle numbers.

Lastly, to detect the onset of superfluidity, we evaluate the superfluid stiffness from the winding number statistics \cite{Ceperley_1995,Pollock_1987}. For a square simulation box with periodic boundary conditions, the superfluid stiffness is given by 
\begin{equation}
    \rho_{s}=\frac{\left\langle W_x^2 + W_y^2\right\rangle}{\beta \mathcal{D}} \,,
\end{equation}
where $W_{\alpha}$ is the winding number along the $\alpha$-axis,
$\beta=(k_BT)^{-1}$ is the inverse temperature and $\mathcal{D}=2$ is the spatial dimension. Since $\rho_s$ is finite only in the superfluid state, it serves as a probe of the BKT transition in our two-dimensional bosonic system. Specifically, the BKT transition temperature, $T_c$, is extracted from the intersection of $\rho_s(T)$ with the universal line $y=\frac{2}{\pi} k_B T$ \cite{nelson_universal_1977}.

\section{Additional Data \label{app:data}}

This section provides additional details and numerical convergence tests that support the findings presented in the main text. We begin with the analytical expression for the chemical potential of the classical dipolar gas. In the high-temperature, dilute limit, this expression is obtained by applying the virial expansion to a classical gas interacting via the pairwise dipolar potential $U_{dd}(r)=D/r^3$. The first-order correction to the non-interacting limit corresponds to $\mu_{dd}$, and is given by \cite{Pathria_statmech_2011,laikhtman_exciton_2009}
\begin{equation}
    \mu_{dd} = \pi \Gamma(1/3) D^{2/3} (k_B T)^{1/3} n \,.
\end{equation}
Comparing this chemical potential with the interaction energy of the classical gas stated in the main text, $\epsilon^{\text{CG}}_{dd} = 2\pi \Gamma(4/3) D^{2/3} (k_BT)^{1/3} n$, we note that they satisfy the proportionality relation $\mu_{dd} = \frac{3}{2} \epsilon^{\text{CG}}_{dd}$.

\begin{figure*}[th!]
\includegraphics[width=0.45\textwidth]{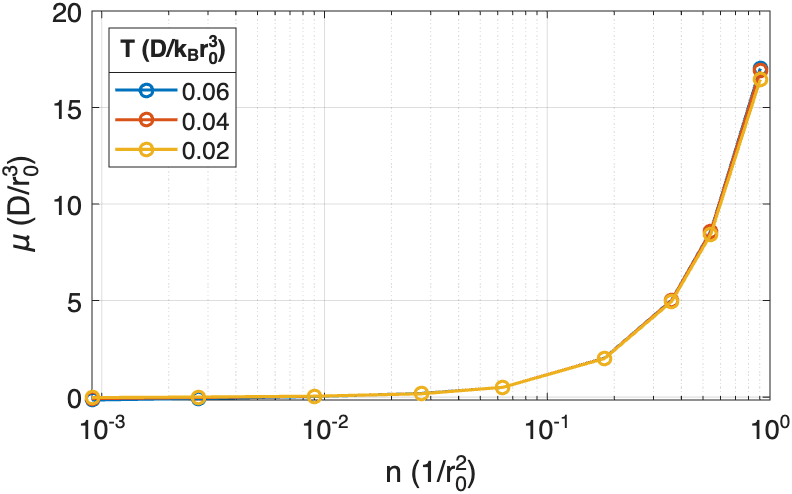}
    \centering
    \caption{Low temperature convergence of $\mu$ at the sampled densities, computed for a system with $N=16$ particles.} \label{fig:mu_conv_low_T}
\end{figure*}

Moving beyond the classical limit, we now test our key assumption that, at low temperatures and high densities, the chemical potential can be estimated from the derivative of the internal energy due to the suppression of thermal fluctuations in these regimes. To validate this assumption, we first verify that our representative low-temperature limit, $T_l$, reflects the effective zero-temperature limit across the considered density range. For that, we compute $\mu$ as a function of density for three decreasing temperatures, including $T_l=0.02 \frac{D}{k_B r_0^3}$. As shown in \cref{fig:mu_conv_low_T}, $\mu$ is well-converged for these temperatures at all considered densities, confirming that $T_l$ indeed represents the ground-state limit of the model, where $\mu$ is determined by the internal energy derivative.

\begin{figure*}[th!]
\includegraphics[width=0.4\textwidth]{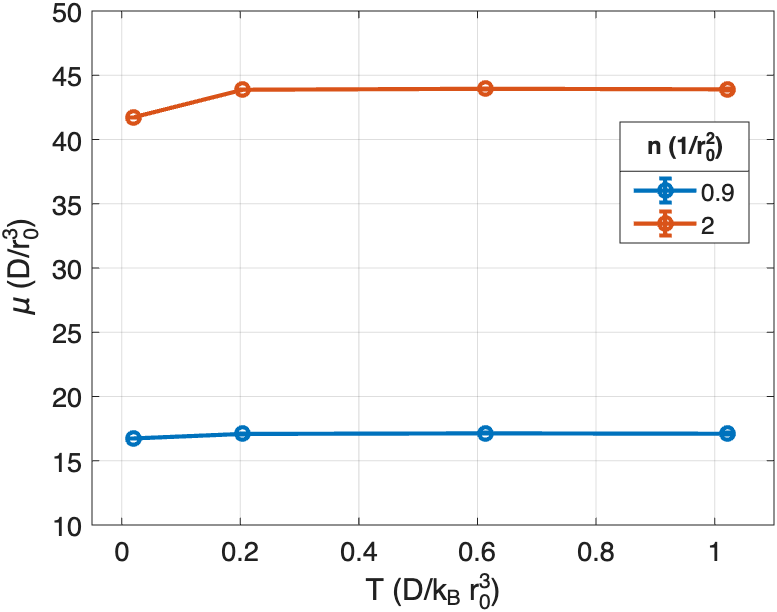}
    \centering
    \caption{Temperature convergence of the chemical potential at the two highest densities $n=0.9, 2$~($1/r_0^2$), computed for a system with $N=16$ particles.} \label{fig:mu_conv_high_n}
\end{figure*}

Furthermore, to justify the use of the internal derivative assumption in the high-density liquid regime at higher temperatures, we verify that $\mu$ is indeed temperature-converged in this limit. In \cref{fig:mu_conv_high_n}, we show the chemical potential as a function of temperature at the two highest simulated densities representing the liquid regime (see \cref{fig:phase_map}). We find that the chemical potential exhibits flat behavior, indicating that thermal effects are negligible in this high-density limit. This observation justifies the use of the energy derivative approach for $\mu$ in the liquid state across both isotherms.

\begin{figure*}[th!]
\includegraphics[width=0.9\textwidth]{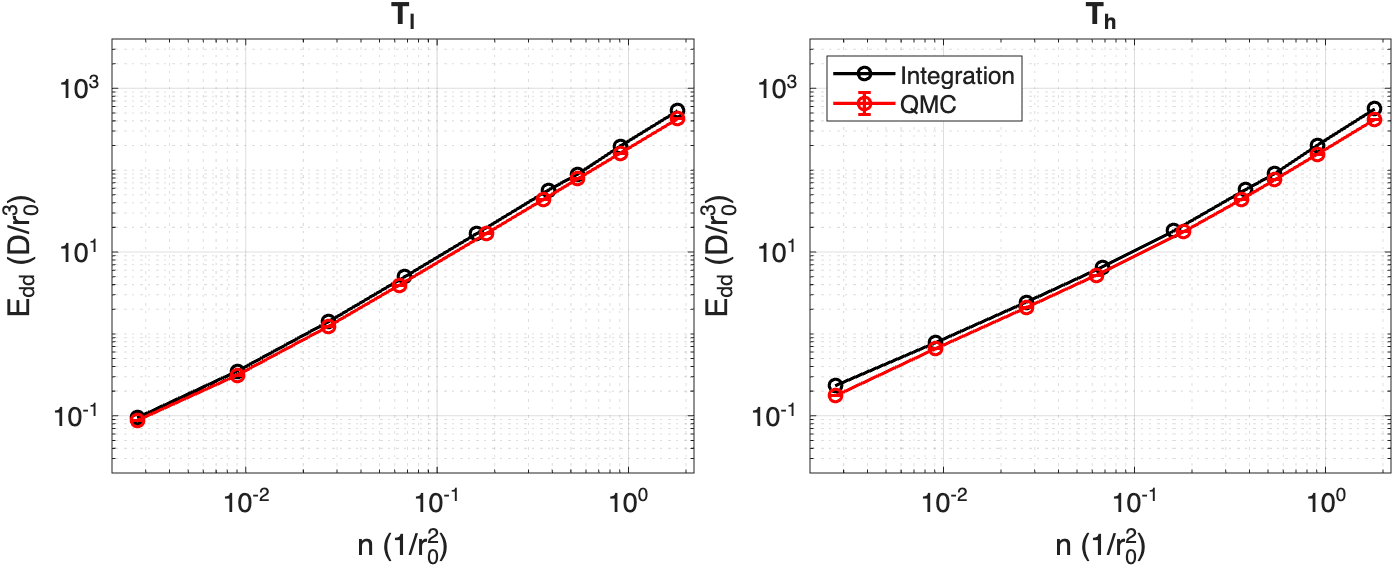}
    \centering
    \caption{The total interaction energy of the dipolar system, obtained via the integration method of \cref{eq:delta_E_def} (black) and directly by QMC (red). The results are shown for a system with $N=32$ particles.} \label{fig:Edd_comp}
\end{figure*}

Next, we verify that the spatial integral definition in \cref{eq:delta_E_def} accurately captures the interaction energy. To this end, we numerically measure the total interaction energy of the system directly via QMC and compare it to the values of $E_{dd}=\frac{N}{2}\epsilon_{dd}$ obtained by integrating the numerically sampled correlation functions $g_2(r)$. The results in \cref{fig:Edd_comp} show excellent agreement, validating the integral approach.

\begin{figure*}[th!]
\includegraphics[width=0.9\textwidth]{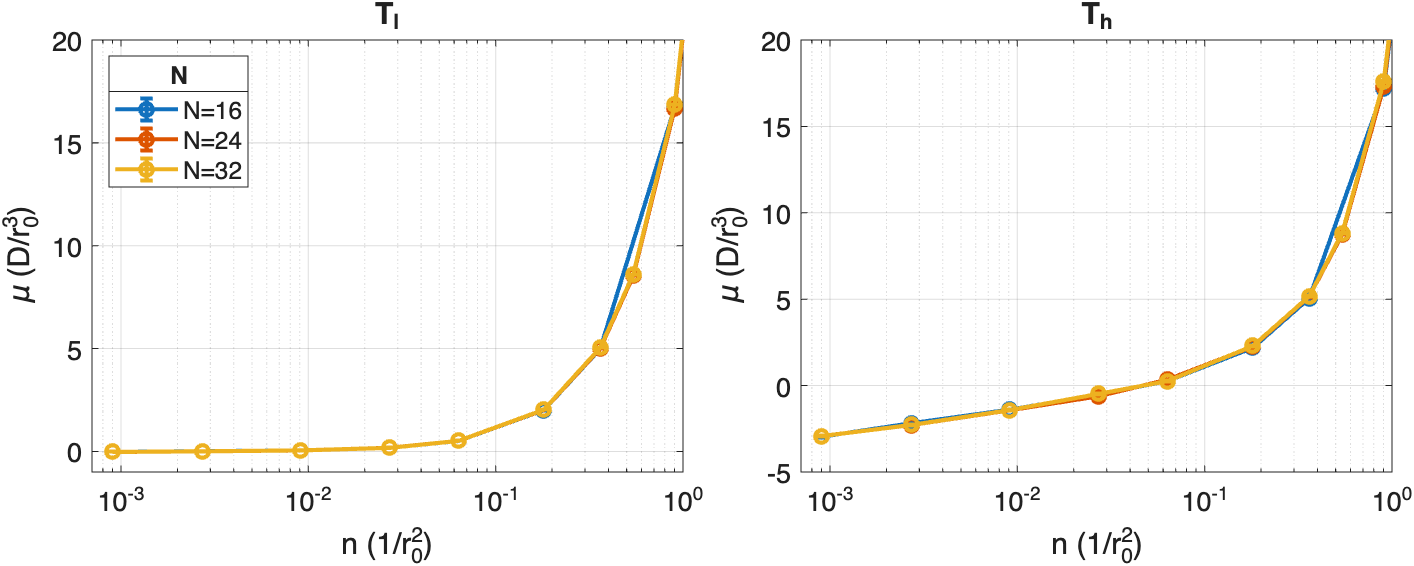}
    \centering
    \caption{System size convergence of the chemical potential, shown for (a) $T_l$ and (b) $T_h$ at the sampled densities.} \label{fig:mu_conv_N}
\end{figure*}

To ensure our numerical data represents the thermodynamic limit, we test the convergence of our results with system size. \cref{fig:mu_conv_N} compares chemical potential curves as a function of density at the two studied isotherms, $T_l$ and $T_h$, calculated for three system sizes. All curves show excellent agreement, with relative errors typically below $1\%$ for $T_l$ and $5\%$ for $T_h$ (away from the crossing point with $\mu=0$). These results corroborate the use of $N=32$ as an effective thermodynamic limit for results presented in this manuscript.

\begin{figure*}[th!]
\includegraphics[width=0.5\textwidth]{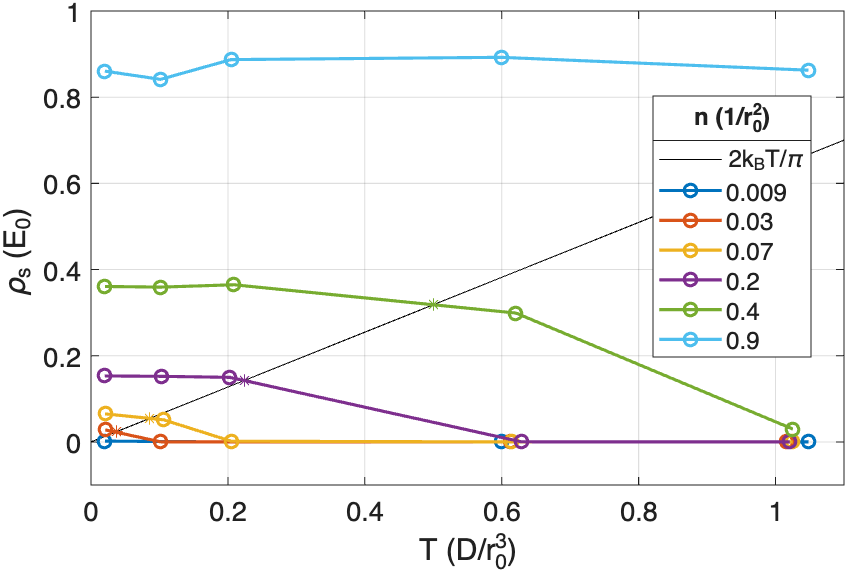}
    \centering
    \caption{Superfluid stiffness as a function of temperature, plotted for different densities, as indicated by the figure legend. The black line is the linear function $y=\frac{2}{\pi}k_B T$, shown in units of $E_0$, and its crossing points with the different $\rho_s$ curves are marked by asterisks. These results were computed for $N=32$ particles.} \label{fig:SF}
\end{figure*}

To further characterize the collective state of the dipolar fluid, we compute $\rho_s$ as a function of temperature for different densities, as shown in \cref{fig:SF}. The BKT transition temperature $T_c$ is then estimated from the intersection of $\rho_s$ with the universal line $y=\frac{2}{\pi}k_B T$ \cite{nelson_universal_1977} and is used to define the shaded superfluid region in the main text figures. Specifically, expecting linear scaling of $T_c$ with density, we use a linear fit to the numerically extracted $T_c$ values to interpolate the transition temperature across the studied density range. We note, however, that definitively determining the BKT transition requires a finite-size scaling analysis of $\rho_s$. Since we extract $T_c$ using data for a single system size, the shaded regions in the main-text figures represent only estimates of the superfluid regime.

\newpage
\bibliography{bib.bib}% Produces the bibliography via BibTeX.

@article{lozovik_strong_2007,
	title = {Strong correlation effects in {2D} {Bose}–{Einstein} condensed dipolar excitons},
	volume = {144},
	issn = {0038-1098},
	url = {http://www.sciencedirect.com/science/article/pii/S0038109807005790},
	doi = {10.1016/j.ssc.2007.07.041},
	number = {9},
	urldate = {2016-10-19},
	journal = {Solid State Communications},
	author = {Lozovik, Yurii E. and Kurbakov, I. L. and Astrakharchik, G. E. and Boronat, J. and Willander, Magnus},
	month = dec,
	year = {2007},
	pages = {399--404}
}

@article{zimmermann_excitonexciton_2007,
	title = {Exciton{\textendash}exciton interaction in coupled quantum wells},
	volume = {144},
	issn = {0038-1098},
	url = {http://www.sciencedirect.com/science/article/pii/S0038109807005832},
	doi = {10.1016/j.ssc.2007.07.044},
	number = {9},
	urldate = {2016-10-19},
	journal = {Solid State Communications},
	author = {Zimmermann, Roland and Schindler, Christoph},
	month = dec,
	year = {2007},
	pages = {395--398}
}

@article{laikhtman_exciton_2009,
	title = {Exciton correlations in coupled quantum wells and their luminescence blue shift},
	volume = {80},
	url = {http://link.aps.org/doi/10.1103/PhysRevB.80.195313},
	doi = {10.1103/PhysRevB.80.195313},
	number = {19},
	urldate = {2015-12-28},
	journal = {Physical Review B},
	author = {Laikhtman, B. and Rapaport, R.},
	month = nov,
	year = {2009},
	pages = {195313},
}

@article{laikhtman_correlations_2009,
	title = {Correlations in a two-dimensional {Bose} gas with long-range interaction},
	volume = {87},
	issn = {0295-5075},
	url = {http://stacks.iop.org/0295-5075/87/i=2/a=27010},
	doi = {10.1209/0295-5075/87/27010},
	number = {2},
	urldate = {2016-10-19},
	journal = {EPL (Europhysics Letters)},
	author = {Laikhtman, B. and Rapaport, R.},
	year = {2009},
	pages = {27010}
}

@article{schindler_analysis_2008,
	title = {Analysis of the exciton-exciton interaction in semiconductor quantum wells},
	volume = {78},
	url = {http://link.aps.org/doi/10.1103/PhysRevB.78.045313},
	doi = {10.1103/PhysRevB.78.045313},
	number = {4},
	urldate = {2016-05-08},
	journal = {Physical Review B},
	author = {Schindler, Christoph and Zimmermann, Roland},
	month = jul,
	year = {2008},
	pages = {045313}
}

@article{nelson_universal_1977,
  title = {Universal Jump in the Superfluid Density of Two-Dimensional Superfluids},
  author = {Nelson, David R. and Kosterlitz, J. M.},
  journal = {Phys. Rev. Lett.},
  volume = {39},
  issue = {19},
  pages = {1201--1205},
  numpages = {0},
  year = {1977},
  month = {Nov},
  publisher = {American Physical Society},
  doi = {10.1103/PhysRevLett.39.1201},
  url = {https://link.aps.org/doi/10.1103/PhysRevLett.39.1201}
}

@article{cohen_dark_2016,
	title = {Dark High Density Dipolar Liquid of Excitons},
	volume = {16},
	issn = {1530-6984},
	url = {http://dx.doi.org/10.1021/acs.nanolett.6b01061},
	doi = {10.1021/acs.nanolett.6b01061},
	number = {6},
	urldate = {2016-06-08},
	journal = {Nano Letters},
	author = {Cohen, Kobi and Shilo, Yehiel and West, Ken and Pfeiffer, Loren and Rapaport, Ronen},
	month = jun,
	year = {2016},
	pages = {3726--3731}
}

@article{zimmerman_collective_2022,
	title = {Collective interlayer pairing and pair superfluidity in vertically stacked layers of dipolar excitons},
	volume = {119},
	issn = {0027-8424, 1091-6490},
	url = {https://pnas.org/doi/full/10.1073/pnas.2205845119},
	doi = {10.1073/pnas.2205845119},
	number = {30},
	urldate = {2023-07-15},
	journal = {Proceedings of the National Academy of Sciences},
	author = {Zimmerman, Michal and Rapaport, Ronen and Gazit, Snir},
	month = jul,
	year = {2022},
	pages = {e2205845119},
}

@Article{Rapaport_2020,
  title = {Attractive interactions, molecular complexes, and polarons in coupled dipolar exciton fluids},
  author = {Hubert, C. and Cohen, K. and Ghazaryan, A. and Lemeshko, M. and Rapaport, R. and Santos, P. V.},
  journal = {Phys. Rev. B},
  volume = {102},
  issue = {4},
  pages = {045307},
  numpages = {12},
  year = {2020},
  month = {Jul},
  publisher = {American Physical Society},
  doi={10.1103/PhysRevB.102.045307}
}

@Article{Fogler_quadrilayer_2021,
  title = {Ground and excited states of coupled exciton liquids in electron-hole quadrilayers},
  author = {Xu, Chao and Fogler, Michael M.},
  journal = {Phys. Rev. B},
  volume = {104},
  issue = {19},
  pages = {195430},
  numpages = {11},
  year = {2021},
  month = {Nov},
  publisher = {American Physical Society},
  doi={10.1103/PhysRevB.104.195430}
}

@article{chomaz_dipolar_2023,
	title = {Dipolar physics: a review of experiments with magnetic quantum gases},
	volume = {86},
	issn = {0034-4885, 1361-6633},
	shorttitle = {Dipolar physics},
	url = {https://iopscience.iop.org/article/10.1088/1361-6633/aca814},
	doi = {10.1088/1361-6633/aca814},
	number = {2},
	urldate = {2023-07-27},
	journal = {Reports on Progress in Physics},
	author = {Chomaz, Lauriane and Ferrier-Barbut, Igor and Ferlaino, Francesca and Laburthe-Tolra, Bruno and Lev, Benjamin L and Pfau, Tilman},
	month = feb,
	year = {2023},
	pages = {026401},
}

@article{hubert_attractive_2019,
	title = {Attractive {Dipolar} {Coupling} between {Stacked} {Exciton} {Fluids}},
	volume = {9},
	issn = {2160-3308},
	url = {https://link.aps.org/doi/10.1103/PhysRevX.9.021026},
	doi = {10.1103/PhysRevX.9.021026},
	number = {2},
	urldate = {2023-07-15},
	journal = {Physical Review X},
	author = {Hubert, Colin and Baruchi, Yifat and Mazuz-Harpaz, Yotam and Cohen, Kobi and Biermann, Klaus and Lemeshko, Mikhail and West, Ken and Pfeiffer, Loren and Rapaport, Ronen and Santos, Paulo},
	month = may,
	year = {2019},
	pages = {021026}
}

@article{shilo_particle_2013,
	title = {Particle correlations and evidence for dark state condensation in a cold dipolar exciton fluid},
	volume = {4},
	copyright = {{\textcopyright} 2013 Nature Publishing Group, a division of Macmillan Publishers Limited. All Rights Reserved.},
	url = {http://www.nature.com/ncomms/2013/130823/ncomms3335/full/ncomms3335.html?message-global=remove},
	doi = {10.1038/ncomms3335},
	urldate = {2015-12-29},
	journal = {Nature Communications},
	author = {Shilo, Y. and Cohen, K. and Laikhtman, B. and West, K. and Pfeiffer, L. and Rapaport, R.},
	month = aug,
	year = {2013},
	pages = {2335}
}

@article{Harpaz_dynamical_2019,
	author = {Yotam Mazuz-Harpaz and Kobi Cohen and Michael Leveson and Ken West and Loren Pfeiffer and Maxim Khodas and Ronen Rapaport},
	doi = {10.1073/pnas.1903374116},
	journal = {Proceedings of the National Academy of Sciences},
	number = {37},
	pages = {18328-18333},
	title = {Dynamical formation of a strongly correlated dark condensate of dipolar excitons},
	url = {https://www.pnas.org/doi/abs/10.1073/pnas.1903374116},
	volume = {116},
	year = {2019},
}

@article{Ceperley_1995,
  title = {Path integrals in the theory of condensed helium},
  author = {Ceperley, D. M.},
  journal = {Rev. Mod. Phys.},
  volume = {67},
  issue = {2},
  pages = {279--355},
  numpages = {0},
  year = {1995},
  month = {Apr},
  publisher = {American Physical Society},
  doi={10.1103/RevModPhys.67.279}
}

@article{Boninsegni_2006,
  title = {Worm algorithm and diagrammatic {M}onte {C}arlo: {A} new approach to continuous-space path integral {M}onte {C}arlo simulations},
  author = {Boninsegni, M. and Prokof'ev, N. V. and Svistunov, B. V.},
  journal = {Phys. Rev. E},
  volume = {74},
  issue = {3},
  pages = {036701},
  numpages = {16},
  year = {2006},
  month = {Sep},
  publisher = {American Physical Society},
  doi={10.1103/PhysRevE.74.036701}
}

@article{Del_Maestro_2014,
  title = {Quantum {M}onte {C}arlo measurement of the chemical potential of ${}^{4}\text{He}$},
  author = {Herdman, C. M. and Rommal, A. and Del Maestro, A.},
  journal = {Phys. Rev. B},
  volume = {89},
  issue = {22},
  pages = {224502},
  numpages = {7},
  year = {2014},
  month = {Jun},
  publisher = {American Physical Society},
  doi={10.1103/PhysRevB.89.224502}
}

@article{Sivalertporn_direct_2012,
  title = {Direct and indirect excitons in semiconductor coupled quantum wells in an applied electric field},
  author = {Sivalertporn, K. and Mouchliadis, L. and Ivanov, A. L. and Philp, R. and Muljarov, E. A.},
  journal = {Phys. Rev. B},
  volume = {85},
  issue = {4},
  pages = {045207},
  numpages = {11},
  year = {2012},
  month = {Jan},
  publisher = {American Physical Society},
  doi={10.1103/PhysRevB.85.045207}
}

@article{Lobanov_theory_2016,
  title = {Theory of condensation of indirect excitons in a trap},
  author = {Lobanov, S. V. and Gippius, N. A. and Butov, L. V.},
  journal = {Phys. Rev. B},
  volume = {94},
  issue = {24},
  pages = {245401},
  numpages = {10},
  year = {2016},
  month = {Dec},
  publisher = {American Physical Society},
  doi={10.1103/PhysRevB.94.245401}
}

@article{Deng_review_2024,
	author = {Eunice Paik and Long Zhang and Kin Fai Mak and Jie Shan and Hui Deng},
	doi = {10.1364/AOP.504035},
	journal = {Adv. Opt. Photon.},
	month = {Dec},
	number = {4},
	pages = {1064--1132},
	publisher = {Optica Publishing Group},
	title = {Excitons and polaritons in two-dimensional transition metal dichalcogenides: a tutorial},
	url = {https://opg.optica.org/aop/abstract.cfm?URI=aop-16-4-1064},
	volume = {16},
	year = {2024},
}

@book{Pathria_statmech_2011,
  title = {Statistical Mechanics},
  author = {Pathria, R. K. and Beale, Paul D.},
  year = {2011},
  edition = {3rd ed},
  publisher = {Elsevier/Academic Press},
  isbn = {978-0-12-382188-1},
  langid = {english},
  lccn = {QC174.8 .P38 2011}
}

@article{Gadway_2016,
	author = {Gadway, Bryce and Yan, Bo},
	doi = {10.1088/0953-4075/49/15/152002},
	journal = {Journal of Physics B: Atomic, Molecular and Optical Physics},
	month = {jun},
	number = {15},
	pages = {152002},
	publisher = {IOP Publishing},
	title = {Strongly interacting ultracold polar molecules},
	url = {https://doi.org/10.1088/0953-4075/49/15/152002},
	volume = {49},
	year = {2016},
}

@article{Baranov_2012,
	author = {Baranov, M. A. and Dalmonte, M. and Pupillo, G. and Zoller, P.},
	doi = {10.1021/cr2003568},
	issn = {0009-2665},
	journal = {Chemical Reviews},
	month = {08},
	number = {9},
	pages = {5012-5061},
	title = {Condensed Matter Theory of Dipolar Quantum Gases},
	url = {https://doi.org/10.1021/cr2003568},
	volume = {112},
	year = {2012},
}

@article{Pollock_1987,
  title = {Path-integral computation of superfluid densities},
  author = {Pollock, E. L. and Ceperley, D. M.},
  journal = {Phys. Rev. B},
  volume = {36},
  issue = {16},
  pages = {8343--8352},
  numpages = {0},
  year = {1987},
  month = {Dec},
  publisher = {American Physical Society},
  doi = {10.1103/PhysRevB.36.8343},
  url = {https://link.aps.org/doi/10.1103/PhysRevB.36.8343}
}

@book{hansen2013theory,
  title     = {Theory of Simple Liquids: With Applications to Soft Matter},
  author    = {Hansen, Jean-Pierre and McDonald, Ian R.},
  edition   = {4th},
  year      = {2013},
  publisher = {Academic Press},
  address   = {Oxford},
  isbn      = {978-0-12-387032-2},
}

@book{chandler1987introduction,
  author    = {Chandler, David},
  title     = {Introduction to Modern Statistical Mechanics},
  publisher = {Oxford University Press},
  year      = {1987},
  address   = {New York},
  isbn      = {978-0195042771}
}

@article{Bottcher2021New,
	author = {B{\"o}ttcher, Fabian and Schmidt, Jan-Niklas and Hertkorn, Jens and Ng, Kevin S H and Graham, Sean D and Guo, Mingyang and Langen, Tim and Pfau, Tilman},
	doi = {10.1088/1361-6633/abc9ab},
    journal = {Reports on Progress in Physics},
	isbn = {0034-4885},
	number = {1},
	pages = {012403},
	publisher = {IOP Publishing},
	title = {New states of matter with fine-tuned interactions: quantum droplets and dipolar supersolids},
	url = {https://doi.org/10.1088/1361-6633/abc9ab},
	volume = {84},
	year = {2021}}

@article{Chomaz2016Quantum,
  title = {Quantum-Fluctuation-Driven Crossover from a Dilute {B}ose-{E}instein {C}ondensate to a {M}acrodroplet in a {D}ipolar {Q}uantum {F}luid},
  author = {Chomaz, L. and Baier, S. and Petter, D. and Mark, M. J. and W{\"a}chtler, F. and Santos, L. and Ferlaino, F.},
  journal = {Phys. Rev. X},
  volume = {6},
  issue = {4},
  pages = {041039},
  numpages = {10},
  year = {2016},
  month = {Nov},
  publisher = {American Physical Society},
  doi = {10.1103/PhysRevX.6.041039},
  url = {https://link.aps.org/doi/10.1103/PhysRevX.6.041039}
}

@article{Ferrier2016Observation,
  title = {Observation of Quantum Droplets in a Strongly Dipolar {B}ose Gas},
  author = {Ferrier-Barbut, Igor and Kadau, Holger and Schmitt, Matthias and Wenzel, Matthias and Pfau, Tilman},
  journal = {Phys. Rev. Lett.},
  volume = {116},
  issue = {21},
  pages = {215301},
  numpages = {6},
  year = {2016},
  month = {May},
  publisher = {American Physical Society},
  doi = {10.1103/PhysRevLett.116.215301},
  url = {https://link.aps.org/doi/10.1103/PhysRevLett.116.215301}
}

@article{Chomaz2019Long,
  title = {Long-Lived and Transient Supersolid Behaviors in Dipolar Quantum Gases},
  author = {Chomaz, L. and Petter, D. and Ilzh{\"o}fer, P. and Natale, G. and Trautmann, A. and Politi, C. and Durastante, G. and van Bijnen, R. M. W. and Patscheider, A. and Sohmen, M. and Mark, M. J. and Ferlaino, F.},
  journal = {Phys. Rev. X},
  volume = {9},
  issue = {2},
  pages = {021012},
  numpages = {12},
  year = {2019},
  month = {Apr},
  publisher = {American Physical Society},
  doi = {10.1103/PhysRevX.9.021012},
  url = {https://link.aps.org/doi/10.1103/PhysRevX.9.021012}
}

@article{Bottcher2019Transient,
  title = {Transient Supersolid Properties in an Array of Dipolar Quantum Droplets},
  author = {B{\"o}ttcher, Fabian and Schmidt, Jan-Niklas and Wenzel, Matthias and Hertkorn, Jens and Guo, Mingyang and Langen, Tim and Pfau, Tilman},
  journal = {Phys. Rev. X},
  volume = {9},
  issue = {1},
  pages = {011051},
  numpages = {7},
  year = {2019},
  month = {Mar},
  publisher = {American Physical Society},
  doi = {10.1103/PhysRevX.9.011051},
  url = {https://link.aps.org/doi/10.1103/PhysRevX.9.011051}
}

@article{Tanzi2019Observation,
  title = {Observation of a Dipolar Quantum Gas with Metastable Supersolid Properties},
  author = {Tanzi, L. and Lucioni, E. and Fam{\`a}, F. and Catani, J. and Fioretti, A. and Gabbanini, C. and Bisset, R. N. and Santos, L. and Modugno, G.},
  journal = {Phys. Rev. Lett.},
  volume = {122},
  issue = {13},
  pages = {130405},
  numpages = {6},
  year = {2019},
  month = {Apr},
  publisher = {American Physical Society},
  doi = {10.1103/PhysRevLett.122.130405},
  url = {https://link.aps.org/doi/10.1103/PhysRevLett.122.130405}
}

@article{Casotti2024Observation,
	author = {Casotti, Eva and Poli, Elena and Klaus, Lauritz and Litvinov, Andrea and Ulm, Clemens and Politi, Claudia and Mark, Manfred J. and Bland, Thomas and Ferlaino, Francesca},
	doi = {10.1038/s41586-024-08149-7},
	id = {Casotti2024},
	isbn = {1476-4687},
	journal = {Nature},
	number = {8038},
	pages = {327--331},
	title = {Observation of vortices in a dipolar supersolid},
	ty = {JOUR},
	url = {https://doi.org/10.1038/s41586-024-08149-7},
	volume = {635},
	year = {2024}}

@article{Biagioni2024Measurements,
	author = {Biagioni, G. and Antolini, N. and Donelli, B. and Pezz{\`e}, L. and Smerzi, A. and Fattori, M. and Fioretti, A. and Gabbanini, C. and Inguscio, M. and Tanzi, L. and Modugno, G.},
	da = {2024/05/01},
	doi = {10.1038/s41586-024-07361-9},
	id = {Biagioni2024},
	isbn = {1476-4687},
	journal = {Nature},
	number = {8013},
	pages = {773--777},
	title = {Measurement of the superfluid fraction of a supersolid by Josephson effect},
	ty = {JOUR},
	url = {https://doi.org/10.1038/s41586-024-07361-9},
	volume = {629},
	year = {2024}}

@Article{Li_2023,
author = {Li, Weijie and Hadjri, Zach and Devenica, Luka M. and Zhang, Jin and Liu, Song and Hone, James and Watanabe, Kenji and Taniguchi, Takashi and Rubio, Angel and Srivastava, Ajit},
id = {Li2023},
isbn = {1476-4660},
journal = {Nature Materials},
number = {12},
pages = {1478--1484},
title = {Quadrupolar--dipolar excitonic transition in a tunnel-coupled van der {W}aals heterotrilayer},
volume = {22},
year = {2023},
doi={10.1038/s41563-023-01667-1}
}

@Article{Xie_2023,
  title = {Bright and Dark Quadrupolar Excitons in the {W}{S}e$_{2}$/{M}o{S}e$_{2}$/{W}{S}e$_{2}$ Heterotrilayer},
  author = {Xie, Yongzhi and Gao, Yuchen and Chen, Fengyu and Wang, Yunkun and Mao, Jun and Liu, Qinyun and Chu, Saisai and Yang, Hong and Ye, Yu and Gong, Qihuang and Feng, Ji and Gao, Yunan},
  journal = {Phys. Rev. Lett.},
  volume = {131},
  issue = {18},
  pages = {186901},
  numpages = {6},
  year = {2023},
  month = {Oct},
  publisher = {American Physical Society},
  doi={10.1103/PhysRevLett.131.186901}
}

@Article{Lian_2023,
author = {Lian, Zhen and Chen, Dongxue and Ma, Lei and Meng, Yuze and Su, Ying and Yan, Li and Huang, Xiong and Wu, Qiran and Chen, Xinyue and Blei, Mark and Taniguchi, Takashi and Watanabe, Kenji and Tongay, Sefaattin and Zhang, Chuanwei and Cui, Yong-Tao and Shi, Su-Fei},
isbn = {2041-1723},
journal = {Nature Communications},
number = {1},
pages = {4604},
title = {Quadrupolar excitons and hybridized interlayer {M}ott insulator in a trilayer moir{\'e} superlattice},
volume = {14},
year = {2023},
doi={10.1038/s41467-023-40288-9},
}

@Article{Xie_2024,
  title = {Transition between quadrupole and staggered dipole interlayer excitons in {W}{S}e$_{2}$/{M}o{S}e$_{2}$/{W}{S}e$_{2}$ heterotrilayers},
  author = {Xie, Yongzhi and Chen, Fengyu and Gao, Yuchen and Wang, Yunkun and Mao, Jun and Liu, Qinyun and Chu, Saisai and Yang, Hong and Ye, Yu and Gong, Qihuang and Feng, Ji and Gao, Yunan},
  journal = {Phys. Rev. B},
  volume = {110},
  issue = {20},
  pages = {L201402},
  numpages = {6},
  year = {2024},
  month = {Nov},
  publisher = {American Physical Society},
  doi={10.1103/PhysRevB.110.L201402}
}

@Article{Bai_2023,
	author = {Bai, Yusong and Li, Yiliu and Liu, Song and Guo, Yinjie and Pack, Jordan and Wang, Jue and Dean, Cory R. and Hone, James and Zhu, Xiaoyang},
	isbn = {1530-6984},
	journal = {Nano Letters},
	journal1 = {Nano Lett.},
	m3 = {doi: 10.1021/acs.nanolett.3c03453},
	month = {12},
	number = {24},
	pages = {11621--11629},
	publisher = {American Chemical Society},
	title = {Evidence for Exciton Crystals in a 2D Semiconductor Heterotrilayer},
	volume = {23},
	year = {2023},
    doi={10.1021/acs.nanolett.3c03453}
}

@Article{Meng_2025,
	author = {Meng, Yuze and Ma, Lei and Yan, Li and Khalifa, Ahmed and Chen, Dongxue and Zhang, Shuai and Banerjee, Rounak and Taniguchi, Takashi and Watanabe, Kenji and Tongay, Seth Ariel and Hunt, Benjamin and Lin, Shi-Zeng and Yao, Wang and Cui, Yong-Tao and Chatterjee, Shubhayu and Shi, Su-Fei},
	isbn = {1749-4893},
	journal = {Nature Photonics},
	number = {11},
	pages = {1219--1224},
	title = {Strong-interaction-driven quadrupolar-to-dipolar exciton transitions in a trilayer moir{\'e} superlattice},
	volume = {19},
	year = {2025},
    doi={10.1038/s41566-025-01741-x}
}

@Article{Slobodkin_2020,
  title = {Quantum Phase Transitions of Trilayer Excitons in Atomically Thin Heterostructures},
  author = {Slobodkin, Yevgeny and Mazuz-Harpaz, Yotam and Refaely-Abramson, Sivan and Gazit, Snir and Steinberg, Hadar and Rapaport, Ronen},
  journal = {Phys. Rev. Lett.},
  volume = {125},
  issue = {25},
  pages = {255301},
  numpages = {7},
  year = {2020},
  month = {Dec},
  publisher = {American Physical Society},
  doi={10.1103/PhysRevLett.125.255301}
}

@article{Zimmerman2026Quadrupolar,
  title = {Quadrupolar and dipolar phases of excitons in transition-metal dichalcogenide trilayer heterostructures},
  author = {Zimmerman, Michal and Podolsky, Daniel and Rapaport, Ronen and Gazit, Snir},
  journal = {Phys. Rev. B},
  volume = {114},
  issue = {18},
  pages = {185116},
  numpages = {22},
  year = {2026},
  month = {Sep},
  publisher = {American Physical Society},
  doi = {10.1103/lz1t-fpyk},
}

@article{Su2023Dipolar,
	author = {Su, Lin and Douglas, Alexander and Szurek, Michal and Groth, Robin and Ozturk, S. Furkan and Krahn, Aaron and H{\'e}bert, Anne H. and Phelps, Gregory A. and Ebadi, Sepehr and Dickerson, Susannah and Ferlaino, Francesca and Markovi{\'c}, Ognjen and Greiner, Markus},
	doi = {10.1038/s41586-023-06614-3},
	id = {Su2023},
	isbn = {1476-4687},
	journal = {Nature},
	number = {7984},
	pages = {724--729},
	title = {Dipolar quantum solids emerging in a {H}ubbard quantum simulator},
	volume = {622},
	year = {2023},
}

@article{Cornish2024Quantum,
	author = {Cornish, Simon L. and Tarbutt, Michael R. and Hazzard, Kaden R. A.},
	da = {2024/05/01},
	doi = {10.1038/s41567-024-02453-9},
	id = {Cornish2024},
	isbn = {1745-2481},
	journal = {Nature Physics},
	number = {5},
	pages = {730--740},
	title = {Quantum computation and quantum simulation with ultracold molecules},
	ty = {JOUR},
	url = {https://doi.org/10.1038/s41567-024-02453-9},
	volume = {20},
	year = {2024}}

@article{Baier2026Extended,
	author = {Baier, S. and Mark, M. J. and Petter, D. and Aikawa, K. and Chomaz, L. and Cai, Z. and Baranov, M. and Zoller, P. and Ferlaino, F.},
	doi = {10.1126/science.aac9812},
	journal = {Science},
	journal1 = {Science},
	month = {2026/10/17},
	number = {6282},
	pages = {201--205},
	publisher = {American Association for the Advancement of Science},
	title = {Extended {B}ose-{H}ubbard models with ultracold magnetic atoms},
	ty = {JOUR},
	url = {https://doi.org/10.1126/science.aac9812},
	volume = {352},
	year = {2016},
	year1 = {2016}}

@article{Lagoin2026Bose,
	author = {Lagoin, Camille and Morin, Corentin and Baldwin, Kirk and Pfeiffer, Loren and Dubin, Fran{\c c}ois},
	da = {2026/04/01},
	doi = {10.1038/s41567-026-03213-7},
	id = {Lagoin2026},
	isbn = {1745-2481},
	journal = {Nature Physics},
	number = {4},
	pages = {566--570},
	title = {Bose--{H}ubbard simulator with long-range hopping},
	ty = {JOUR},
	url = {https://doi.org/10.1038/s41567-026-03213-7},
	volume = {22},
	year = {2026}}

@article{Lagoin2022Extended,
	author = {Lagoin, C. and Bhattacharya, U. and Grass, T. and Chhajlany, R. W. and Salamon, T. and Baldwin, K. and Pfeiffer, L. and Lewenstein, M. and Holzmann, M. and Dubin, F.},
	da = {2022/09/01},
	doi = {10.1038/s41586-022-05123-z},
	id = {Lagoin2022},
	isbn = {1476-4687},
	journal = {Nature},
	number = {7927},
	pages = {485--489},
	title = {Extended {B}ose--{H}ubbard model with dipolar excitons},
	ty = {JOUR},
	url = {https://doi.org/10.1038/s41586-022-05123-z},
	volume = {609},
	year = {2022}}

@article{Buchler2007Strongly,
  title = {Strongly Correlated 2{D} Quantum Phases with Cold Polar Molecules: Controlling the Shape of the Interaction Potential},
  author = {B{\"u}chler, H. P. and Demler, E. and Lukin, M. and Micheli, A. and Prokof'ev, N. and Pupillo, G. and Zoller, P.},
  journal = {Phys. Rev. Lett.},
  volume = {98},
  issue = {6},
  pages = {060404},
  numpages = {4},
  year = {2007},
  publisher = {American Physical Society},
  doi = {10.1103/PhysRevLett.98.060404},
  url = {https://link.aps.org/doi/10.1103/PhysRevLett.98.060404}
}

@article{Astrakharchik2007Quantum,
  title = {Quantum Phase Transition in a Two-Dimensional System of Dipoles},
  author = {Astrakharchik, G. E. and Boronat, J. and Kurbakov, I. L. and Lozovik, Yu. E.},
  journal = {Phys. Rev. Lett.},
  volume = {98},
  issue = {6},
  pages = {060405},
  numpages = {4},
  year = {2007},
  publisher = {American Physical Society},
  doi = {10.1103/PhysRevLett.98.060405},
  url = {https://link.aps.org/doi/10.1103/PhysRevLett.98.060405}
}

@article{Combescot_2007,
  title = {Bose-{E}instein Condensation in Semiconductors: The Key Role of Dark Excitons},
  author = {Combescot, Monique and Betbeder-Matibet, Odile and Combescot, Roland},
  journal = {Phys. Rev. Lett.},
  volume = {99},
  issue = {17},
  pages = {176403},
  numpages = {4},
  year = {2007},
  publisher = {American Physical Society},
  doi={10.1103/PhysRevLett.99.176403}
}

@article{Beian_2017,
	author = {Beian, Mussie and Alloing, Mathieu and Anankine, Romain and Cambril, Edmond and Gomez Carbonell, Carmen and Lema{\^\i}tre, Aristide and Dubin, Fran{\c c}ois},
	journal = {Europhysics Letters},
	doi = {10.1209/0295-5075/119/37004},
	isbn = {0295-5075},
	number = {3},
	pages = {37004},
	publisher = {EDP Sciences, IOP Publishing and Societ{\`a}Italiana di Fisica},
	title = {Spectroscopic signatures for the dark {B}ose-{E}instein condensation of spatially indirect excitons},
	url = {https://doi.org/10.1209/0295-5075/119/37004},
	volume = {119},
	year = {2017},
}

@article{Misra_2022,
	author = {Misra, Subhradeep and Stern, Michael and Umansky, Vladimir and Bar-Joseph, Israel},
	doi = {10.1073/pnas.2203531119},
	journal = {Proceedings of the National Academy of Sciences},
	number = {32},
	pages = {e2203531119},
	title = {The Role of Spin-Flip Collisions in a Dark-Exciton Condensate},
	url = {https://doi.org/10.1073/pnas.2203531119},
	volume = {119},
	year = {2022},
}

@article{Remeika2015Measurements,
  title = {Measurement of exciton correlations using electrostatic lattices},
  author = {Remeika, M. and Leonard, J. R. and Dorow, C. J. and Fogler, M. M. and Butov, L. V. and Hanson, M. and Gossard, A. C.},
  journal = {Phys. Rev. B},
  volume = {92},
  issue = {11},
  pages = {115311},
  numpages = {10},
  year = {2015},
  month = {Sep},
  publisher = {American Physical Society},
  doi = {10.1103/PhysRevB.92.115311},
  url = {https://link.aps.org/doi/10.1103/PhysRevB.92.115311}
}

@article{Sun2022Excitonic,
	author = {Sun, Zhe and Ciarrocchi, Alberto and Tagarelli, Fedele and Gonzalez Marin, Juan Francisco and Watanabe, Kenji and Taniguchi, Takashi and Kis, Andras},
	doi = {10.1038/s41566-021-00908-6},
	id = {Sun2022},
	isbn = {1749-4893},
	journal = {Nature Photonics},
	number = {1},
	pages = {79--85},
	title = {Excitonic transport driven by repulsive dipolar interaction in a van der {W}aals heterostructure},
	url = {https://doi.org/10.1038/s41566-021-00908-6},
	volume = {16},
	year = {2022}}

@article{Schinner2013Manybody,
  title = {Many-body correlations of electrostatically trapped dipolar excitons},
  author = {Schinner, G. J. and Repp, J. and Schubert, E. and Rai, A. K. and Reuter, D. and Wieck, A. D. and Govorov, A. O. and Holleitner, A. W. and Kotthaus, J. P.},
  journal = {Phys. Rev. B},
  volume = {87},
  issue = {20},
  pages = {205302},
  numpages = {8},
  year = {2013},
  month = {May},
  publisher = {American Physical Society},
  doi = {10.1103/PhysRevB.87.205302},
  url = {https://link.aps.org/doi/10.1103/PhysRevB.87.205302}
}

@article{Remeika_2009_localization,
  title = {Localization-Delocalization Transition of Indirect Excitons in Lateral Electrostatic Lattices},
  author = {Remeika, M. and Graves, J. C. and Hammack, A. T. and Meyertholen, A. D. and Fogler, M. M. and Butov, L. V. and Hanson, M. and Gossard, A. C.},
  journal = {Phys. Rev. Lett.},
  volume = {102},
  issue = {18},
  pages = {186803},
  numpages = {4},
  year = {2009},
  publisher = {American Physical Society},
  doi={10.1103/PhysRevLett.102.186803}
}

@article{Alloing_2014_evidence,
	author = {Alloing, Mathieu and Beian, Mussie and Lewenstein, Maciej and Fuster, David and Gonz{\'a}lez, Yolanda and Gonz{\'a}lez, Luisa and Combescot, Roland and Combescot, Monique and Dubin, Fran{\c c}ois},
	journal = {Europhysics Letters},
	number = {1},
	pages = {10012},
	publisher = {EDP Sciences, IOP Publishing and Societ{\`a} Italiana di Fisica},
	title = {Evidence for a {B}ose-{E}instein condensate of excitons},
	volume = {107},
	year = {2014},
    doi={ 10.1209/0295-5075/107/10012}
}

@article{Anankine_2017,
  title = {Quantized Vortices and Four-Component Superfluidity of Semiconductor Excitons},
  author = {Anankine, Romain and Beian, Mussie and Dang, Suzanne and Alloing, Mathieu and Cambril, Edmond and Merghem, Kamel and Carbonell, Carmen Gomez and Lema{\^{i}}tre, Aristide and Dubin, Fran\ifmmode \mbox{\c{c}}\else \c{c}\fi{}ois},
  journal = {Phys. Rev. Lett.},
  volume = {118},
  issue = {12},
  pages = {127402},
  numpages = {5},
  year = {2017},
  publisher = {American Physical Society},
  doi={10.1103/PhysRevLett.118.127402}
}

@article{Stern2008Photoluminescence,
  title = {Photoluminescence Ring Formation in Coupled Quantum Wells: Excitonic Versus Ambipolar Diffusion},
  author = {Stern, M. and Garmider, V. and Segre, E. and Rappaport, M. and Umansky, V. and Levinson, Y. and Bar-Joseph, I.},
  journal = {Phys. Rev. Lett.},
  volume = {101},
  issue = {25},
  pages = {257402},
  numpages = {4},
  year = {2008},
  month = {Dec},
  publisher = {American Physical Society},
  doi = {10.1103/PhysRevLett.101.257402},
  url = {https://link.aps.org/doi/10.1103/PhysRevLett.101.257402}
}

@article{MacDonald2018Theory,
  title = {Theory of optical absorption by interlayer excitons in transition metal dichalcogenide heterobilayers},
  author = {Wu, Fengcheng and Lovorn, Timothy and MacDonald, A. H.},
  journal = {Phys. Rev. B},
  volume = {97},
  issue = {3},
  pages = {035306},
  numpages = {10},
  year = {2018},
  month = {Jan},
  publisher = {American Physical Society},
  doi = {10.1103/PhysRevB.97.035306},
  url = {https://link.aps.org/doi/10.1103/PhysRevB.97.035306}
}

@Article{Goryca_2019_revealing,
	author = {Goryca, M. and Li, J. and Stier, A. V. and Taniguchi, T. and Watanabe, K. and Courtade, E. and Shree, S. and Robert, C. and Urbaszek, B. and Marie, X. and Crooker, S. A.},
	journal = {Nature Communications},
	number = {1},
	pages = {4172},
	title = {Revealing exciton masses and dielectric properties of monolayer semiconductors with high magnetic fields},
	volume = {10},
	year = {2019},
    doi={10.1038/s41467-019-12180-y}
}

@book{levinshtein1999handbook,
  title={Handbook Series on Semiconductor Parameters: Ternary and quaternary III-V compounds},
  author={Levinshtein, M. and Rumyantsev, S.L. and Shur, M. and Scientific, W.},
  year={1999},
  publisher={World Scientific Publishing Company},
}

@Article{Mak2021EI,
	author = {Ma, Liguo and Nguyen, Phuong X. and Wang, Zefang and Zeng, Yongxin and Watanabe, Kenji and Taniguchi, Takashi and MacDonald, Allan H. and Mak, Kin Fai and Shan, Jie},
	id = {Ma2021},
	isbn = {1476-4687},
	journal = {Nature},
	number = {7882},
	pages = {585--589},
	title = {Strongly correlated excitonic insulator in atomic double layers},
	volume = {598},
	year = {2021},
    doi={10.1038/s41586-021-03947-9}

}

@Article{Mak2022EI,
	author = {Gu, Jie and Ma, Liguo and Liu, Song and Watanabe, Kenji and Taniguchi, Takashi and Hone, James C. and Shan, Jie and Mak, Kin Fai},
	isbn = {1745-2481},
	journal = {Nature Physics},
	number = {4},
	pages = {395--400},
	title = {Dipolar excitonic insulator in a moir{\'e} lattice},
	volume = {18},
	year = {2022},
    doi={10.1038/s41567-022-01532-z}
}

@Article{Watanabe2022EI,
	author = {Zhang, Zuocheng and Regan, Emma C. and Wang, Danqing and Zhao, Wenyu and Wang, Shaoxin and Sayyad, Mohammed and Yumigeta, Kentaro and Watanabe, Kenji and Taniguchi, Takashi and Tongay, Sefaattin and Crommie, Michael and Zettl, Alex and Zaletel, Michael P. and Wang, Feng},
	id = {Zhang2022},
	isbn = {1745-2481},
	journal = {Nature Physics},
	number = {10},
	pages = {1214--1220},
	title = {Correlated interlayer exciton insulator in heterostructures of monolayer {W}{S}e$_2$ and moir{\'e} {W}{S}$_2$/{W}{S}e$_2$},
	volume = {18},
	year = {2022},
    doi={10.1038/s41567-022-01702-z}
}

@Article{Watanabe2020Mott,
	author = {Tang, Yanhao and Li, Lizhong and Li, Tingxin and Xu, Yang and Liu, Song and Barmak, Katayun and Watanabe, Kenji and Taniguchi, Takashi and MacDonald, Allan H. and Shan, Jie and Mak, Kin Fai},
	isbn = {1476-4687},
	journal = {Nature},
	number = {7799},
	pages = {353--358},
	title = {Simulation of {H}ubbard model physics in {W}{S}e$_2$/{W}{S}$_2$ moir{\'e} superlattices},
	volume = {579},
	year = {2020},
    doi={s41586-020-2085-3}
}

@Article{Mak2020Mott,
	author = {Xu, Yang and Liu, Song and Rhodes, Daniel A. and Watanabe, Kenji and Taniguchi, Takashi and Hone, James and Elser, Veit and Mak, Kin Fai and Shan, Jie},
	isbn = {1476-4687},
	journal = {Nature},
	number = {7833},
	pages = {214--218},
	title = {Correlated insulating states at fractional fillings of moir{\'e} superlattices},
	volume = {587},
	year = {2020},
    doi={10.1038/s41586-020-2868-6}
}

@Article{Imamoglu2020Mott,
	author = {Shimazaki, Yuya and Schwartz, Ido and Watanabe, Kenji and Taniguchi, Takashi and Kroner, Martin and Imamo{\u g}lu, Ata{\c c}},
	id = {Shimazaki2020},
	isbn = {1476-4687},
	journal = {Nature},
	number = {7804},
	pages = {472--477},
	title = {Strongly correlated electrons and hybrid excitons in a moir{\'e} heterostructure},
	volume = {580},
	year = {2020},
    doi={10.1038/s41586-020-2191-2}
}

@Article{Wang2020Mott,
	author = {Wang, Lei and Shih, En-Min and Ghiotto, Augusto and Xian, Lede and Rhodes, Daniel A. and Tan, Cheng and Claassen, Martin and Kennes, Dante M. and Bai, Yusong and Kim, Bumho and Watanabe, Kenji and Taniguchi, Takashi and Zhu, Xiaoyang and Hone, James and Rubio, Angel and Pasupathy, Abhay N. and Dean, Cory R.},
	isbn = {1476-4660},
	journal = {Nature Materials},
	number = {8},
	pages = {861--866},
	title = {Correlated electronic phases in twisted bilayer transition metal dichalcogenides},
	volume = {19},
	year = {2020},
    doi={10.1038/s41563-020-0708-6}
}

@article{Gao_2024_excitonic,
	author = {Gao, Beini and Su{\'a}rez-Forero, Daniel G. and Sarkar, Supratik and Huang, Tsung-Sheng and Session, Deric and Mehrabad, Mahmoud Jalali and Ni, Ruihao and Xie, Ming and Upadhyay, Pranshoo and Vannucci, Jonathan and Mittal, Sunil and Watanabe, Kenji and Taniguchi, Takashi and Imamoglu, Atac and Zhou, You and Hafezi, Mohammad},
	doi = {10.1038/s41467-024-46616-x},
	journal = {Nature Communications},
	number = {1},
	pages = {2305},
	title = {Excitonic {M}ott insulator in a {B}ose-{F}ermi-Hubbard system of moir{\'e} {W}{S}$_2$/{W}{S}e$_2$ heterobilayer},
	volume = {15},
	year = {2024},
}

@article{Seyler_2019_signatures,
	author = {Seyler, Kyle L. and Rivera, Pasqual and Yu, Hongyi and Wilson, Nathan P. and Ray, Essance L. and Mandrus, David G. and Yan, Jiaqiang and Yao, Wang and Xu, Xiaodong},
	doi = {10.1038/s41586-019-0957-1},
	journal = {Nature},
	number = {7746},
	pages = {66--70},
	title = {Signatures of moir{\'e}-trapped valley excitons in {M}o{S}e$_2$/{W}{S}e$_2$ heterobilayers},
	volume = {567},
	year = {2019},
}

\end{document}